\documentclass[10pt,journal,compsoc]{IEEEtran}
\usepackage{acro}
\usepackage[utf8]{inputenc}
\usepackage[T1]{fontenc}
\usepackage{textcomp}
\usepackage{fancyvrb}
\usepackage{algpseudocode}
\usepackage{mathtools}
\usepackage{bbm}
\usepackage[flushleft]{threeparttable}	
\usepackage{etoolbox}\AtBeginEnvironment{algorithmic}{\scriptsize} 
\usepackage{adjustbox}
\usepackage{pgfplots}
\usepackage{cancel}
\usepackage{pgfplotstable}
\usepackage{filecontents}
\usepackage{dblfloatfix}    
\usepackage{multirow}
\usepgfplotslibrary{colorbrewer}
\usetikzlibrary{pgfplots.statistics, pgfplots.colorbrewer} \usepackage{forest}
\usetikzlibrary{arrows.meta, shapes.geometric, calc, shadows}
\usetikzlibrary{fit,calc}
\usepackage{smartdiagram}
\usepackage[ruled]{algorithm2e}
\newcommand*{\tikzmk}[1]{\tikz[remember picture,overlay,] \node (#1) {};\ignorespaces}
\newcommand{\boxit}[1]{\tikz[remember picture,overlay]{\node[yshift=3pt,fill=#1,opacity=.25,fit={(A)($(B)+(.95\linewidth,.8\baselineskip)$)}] {};}\ignorespaces}

\colorlet{mygreen}{green!75!black}
\colorlet{col1in}{red!30}
\colorlet{col1out}{red!40}
\colorlet{col2in}{mygreen!40}
\colorlet{col2out}{mygreen!50}
\colorlet{col3in}{blue!30}
\colorlet{col3out}{blue!40}
\colorlet{col4in}{mygreen!20}
\colorlet{col4out}{mygreen!30}
\colorlet{col5in}{blue!10}
\colorlet{col5out}{blue!20}
\colorlet{col6in}{blue!20}
\colorlet{col6out}{blue!30}
\colorlet{col7out}{orange}
\colorlet{col7in}{orange!50}
\colorlet{col8out}{orange!40}
\colorlet{col8in}{orange!20}
\colorlet{linecol}{blue!60}
\usepgfplotslibrary{colorbrewer}
\colorlet{backgroundcol}{yellow!10!white}

\usetikzlibrary{decorations.pathmorphing}
\tikzset{discont/.style={decoration={zigzag,segment length=1.5pt, amplitude=4pt},decorate}}
\def\discontarrow(#1)(#2)(#3)(#4);{
	\draw[discont] (#2) -- (#3);
	\draw[] (#1) -- (#2) (#3) -- (#4);
}

\newcommand{\highlight}[1]{%
	\par\noindent
	\fcolorbox{black}{backgroundcol}{%
		\parbox{\dimexpr\linewidth-2\fboxsep\relax}{%
			#1
		}%
}}

\usepgflibrary{decorations.fractals}
\pgfplotstableset{
rowfont/.style={
	postproc cell content/.append code={
		\count0=\pgfplotstablerow
		\advance\count0 by1
		\ifnum\count0=#1
		\pgfkeysalso{@cell content/.add={\bfseries\boldmath\color{red}}{}}
		\fi
	},
},
colfont/.style={
	postproc cell content/.append code={
		\count0=\pgfplotstablecol
		\advance\count0 by1
		\ifnum\count0=#1
		\pgfkeysalso{@cell content/.add={\bfseries\boldmath\color{red}}{}}
		\fi
	},
},	
colfontwhite/.style={
	postproc cell content/.append code={
		\count0=\pgfplotstablecol
		\advance\count0 by1
		\ifnum\count0=#1
		\pgfkeysalso{@cell content/.add={\bfseries\boldmath\color{white}}{}}
		\fi
	},
},	
}

\usepackage{colortbl}
\usepgfplotslibrary{colormaps}
\pgfplotsset{colormap/Set3} 
\usepackage{cellspace}
\newcolumntype{C}{}
\newlength{\cellspacelimit}
\pgfplotstableset{
	/color cells/min/.initial=0,
	/color cells/max/.initial=100,
	/color cells/textcolor/.initial=,
	color cells/.code={%
		\pgfqkeys{/color cells}{#1}%
		\pgfkeysalso{%
			postproc cell content/.code={%
				\begingroup
				\pgfkeysgetvalue{/pgfplots/table/@preprocessed
					cell content}\value
				\ifx\value\empty
				\endgroup
				\else
				\pgfmathfloatparsenumber{\value}%
				\pgfmathfloattofixed{\pgfmathresult}%
				\let\value=\pgfmathresult
				\pgfplotscolormapaccess
				[\pgfkeysvalueof{/color cells/min}:\pgfkeysvalueof{/color
					cells/max}]
				{\value}
				{\pgfkeysvalueof{/pgfplots/colormap name}}%
				\pgfkeysgetvalue{/pgfplots/table/@cell content}\typesetvalue
				\pgfkeysgetvalue{/color cells/textcolor}\textcolorvalue
				\toks0=\expandafter{\typesetvalue}%
				\xdef\temp{%
					\noexpand\pgfkeysalso{%
						@cell content={%
							\noexpand\cellcolor[rgb]{\pgfmathresult}%
							\noexpand\definecolor{mapped
								color}{rgb}{\pgfmathresult}%
							\ifx\textcolorvalue\empty
							\else
							\noexpand\color{\textcolorvalue}%
							\fi
							\the\toks0 %
						}%
					}%
				}%
				\endgroup
				\temp
				\fi
			}%
		}%
	},
	every column/.style={column type={Cc!{\color{white}\vrule width 1pt}}},
	every first column/.style={reset styles, column type={l}, string type},
	every head row/.style={before row=\toprule, after row=\midrule},
	after row={\noalign{\global\arrayrulewidth=1pt}\arrayrulecolor{white}\hline},
	every last row/.style={after row=\arrayrulecolor{black}\bottomrule}, 
	@content options for rows/.style 2 args={
		postproc cell content/.add code={%
			\def\isInInputList{0}%
			\pgfplotsforeachungrouped \II in {#1} {%
				\ifnum\II<0
				\count0=\II
				\advance\count0 by\pgfplotstablerows
				\edef\II{\the\count0 }%
				\fi
				\ifnum\II=\pgfplotstablerow\relax
				\def\isInInputList{1}%
				\fi
			}%
			\if1\isInInputList%
			\pgfkeysalso{#2}%
			\fi
		}{},
	},
}

\pgfkeys{/pgf/number format/.cd,1000 sep={\,},fixed,precision=3}

\makeatletter
\pgfplotsset{
	boxplot/hide outliers/.code={
		\def\pgfplotsplothandlerboxplot@outlier{}%
	}
}
\makeatother

\newcolumntype{a}{>{\columncolor{Cyan!10!white}}c}
\newcolumntype{b}{>{\columncolor{Gray!10!white}}c}
\newcolumntype{w}{>{\columncolor{red!10!white}}c}
\newcolumntype{x}{>{\columncolor{green!10!white}}c}
\newcolumntype{y}{>{\columncolor{blue!10!white}}c}
\newcolumntype{z}{>{\columncolor{yellow!10!white}}c}
\definecolor{myblue}{RGB}{37,165,203}
\definecolor{myred}{RGB}{175,32,67}

\usepgfplotslibrary{units}
\usetikzlibrary{arrows, calc, spy, patterns, backgrounds,decorations.pathreplacing, arrows.meta}

\pgfplotsset{
	my ybar legend/.style={
		legend image code/.code={
			\draw [##1] (0cm,-0.6ex) rectangle +(1.75em,1.1ex);
		},
	},
}

\makeatletter
\pgfdeclarepatternformonly[\hatchdistance,\hatchthickness]{flexible hatch}
{\pgfqpoint{0pt}{0pt}}
{\pgfqpoint{\hatchdistance}{\hatchdistance}}
{\pgfpoint{\hatchdistance-1pt}{\hatchdistance-1pt}}%
{
	\pgfsetcolor{\tikz@pattern@color}
	\pgfsetlinewidth{\hatchthickness}
	\pgfpathmoveto{\pgfqpoint{0pt}{0pt}}
	\pgfpathlineto{\pgfqpoint{\hatchdistance}{\hatchdistance}}
	\pgfusepath{stroke}
}

\makeatletter
\pgfplotsset{
	discontinuous line/.code={
		\pgfkeysalso{mesh, shorten <=#1, shorten >=#1,
			legend image code/.code={
				\draw [##1, shorten <=0cm] (0cm,0cm) -- (0.3cm,0cm);
				\draw [only marks] plot coordinates {(0.3cm,0cm)};
				\draw [##1, shorten >=0cm] (0.3cm,0cm) -- (0.6cm,0cm);
		}}
		\def\pgfplotsplothandlermesh@VISUALIZE@std@fill@andor@stroke{%
			\pgfplotspatchclass{\pgfplotsplothandlermesh@patchclass}{fill path}%
			\pgfplotsplothandlermesh@definecolor
			\pgfusepath{stroke}
			\pgfplotsplothandlermesh@show@normals@if@configured
		}%
	},
	discontinuous line/.default=1.5mm
}
\makeatother

\usetikzlibrary{patterns}
	\newlength{\hatchspread}
	\newlength{\hatchthickness}
	\newlength{\hatchshift}
	\newcommand{\hatchcolor}{}
	\tikzset{hatchspread/.code={\setlength{\hatchspread}{#1}},
		hatchthickness/.code={\setlength{\hatchthickness}{#1}},
		hatchshift/.code={\setlength{\hatchshift}{#1}},
		hatchcolor/.code={\renewcommand{\hatchcolor}{#1}}}
	\tikzset{hatchspread=3pt,
		hatchthickness=0.4pt,
		hatchshift=0pt,
		hatchcolor=black}
	\pgfdeclarepatternformonly[\hatchspread,\hatchthickness,\hatchshift,\hatchcolor]
	{custom north west lines}
	{\pgfqpoint{\dimexpr-2\hatchthickness}{\dimexpr-2\hatchthickness}}
	{\pgfqpoint{\dimexpr\hatchspread+2\hatchthickness}{\dimexpr\hatchspread+2\hatchthickness}}
	{\pgfqpoint{\dimexpr\hatchspread}{\dimexpr\hatchspread}}
	{
		\pgfsetlinewidth{\hatchthickness}
		\pgfpathmoveto{\pgfqpoint{0pt}{\dimexpr\hatchspread+\hatchshift}}
		\pgfpathlineto{\pgfqpoint{\dimexpr\hatchspread+0.15pt+\hatchshift}{-0.15pt}}
		\ifdim \hatchshift > 0pt
		\pgfpathmoveto{\pgfqpoint{0pt}{\hatchshift}}
		\pgfpathlineto{\pgfqpoint{\dimexpr0.15pt+\hatchshift}{-0.15pt}}
		\fi
		\pgfsetstrokecolor{\hatchcolor}
		\pgfusepath{stroke}
	}
	\pgfdeclarepatternformonly[\hatchspread,\hatchthickness,\hatchshift,\hatchcolor]
	{custom north east lines}
	{\pgfqpoint{\dimexpr-2\hatchthickness}{\dimexpr-2\hatchthickness}}
	{\pgfqpoint{\dimexpr\hatchspread+2\hatchthickness}{\dimexpr\hatchspread+2\hatchthickness}}
	{\pgfqpoint{\dimexpr\hatchspread}{\dimexpr\hatchspread}}
	{
		\pgfsetlinewidth{\hatchthickness}
		\pgfpathmoveto{\pgfqpoint{\dimexpr\hatchshift-0.15pt}{-0.15pt}}
		\pgfpathlineto{\pgfqpoint{\dimexpr\hatchspread+0.15pt}{\dimexpr\hatchspread-\hatchshift+0.15pt}}
		\ifdim \hatchshift > 0pt
		\pgfpathmoveto{\pgfqpoint{-0.15pt}{\dimexpr\hatchspread-\hatchshift-0.15pt}}
		\pgfpathlineto{\pgfqpoint{\dimexpr\hatchshift+0.15pt}{\dimexpr\hatchspread+0.15pt}}
		\fi
		\pgfsetstrokecolor{\hatchcolor}
		\pgfusepath{stroke}
	}

\usepackage{booktabs}
\usepackage{sidecap}
\usepackage{graphicx, caption, subcaption}
\usepackage{relsize}
\usepackage{comment}

\usepackage{amsmath}
\usepackage{amsfonts}
\usepackage{amsthm}
\usepackage{relsize}
\usepackage{xspace}

\usepackage{todonotes}
\usepackage{setspace}
\usepackage{cleveref}
\usepackage[binary-units=true]{siunitx}

\newcommand{\acrodef}[2]{\DeclareAcronym{#1}{short={#1},long={#2}}}
\acrodef{AABB}{axis-aligned bounding box}
\acrodef{API}{application programming interface}
\acrodef{ASIC}{application specific integrated circuit}
\acrodef{AST}{abstract syntax tree}
\acrodef{AVX}{advanced vector extensions}
\acrodef{BRAM}{block RAM}
\acrodef{CB}{compute-bound}
\acrodef{CER}{communication-to-execution ratio}
\acrodef{CG}{conjugate gradient}
\acrodef{ChebFD}{Chebyshev filter diagonalization}
\acrodef{CL}{cache line}
\acrodef{CoD}{cluster-on-die}
\acrodef{CPI}{cycles per instruction}
\acrodef{CPU}{central processing unit}
\acrodef{CUDA}{compute unified device architecture}
\acrodef{CST}{concrete syntax tree}
\acrodef{DP}{double precision}
\acrodef{DPM}{delay propagation mechanism}
\acrodef{DOF}{degree of freedom}
\acrodef{DOFS}{degrees of freedom}
\acrodef{DOM}{delay overlapping mechanism}
\acrodef{DPOM}{delay propagation and overlapping mechanisms}
\acrodef{DSL}{domain-specific language}
\acrodef{DVFS}{dynmic voltage frequency scaling}
\acrodef{ECM}{execution-cache-memory}
\acrodef{FD}{finite difference}
\acrodef{FEM}{finite element method}
\acrodef{FFC}{FEniCS Form Compiler}
\acrodef{FFT}{Fast Fourier transform}
\acrodef{FIFO}{first in first out}
\acrodef{FLOPS}{floating point operations per second}
\acrodef{FMA}{fused multiply-add}
\acrodef{FP}{floating-point}
\acrodef{FPGA}{field-programmable gate array}
\acrodef{FV}{finite volume}
\acrodef{GMRES}{generalized minimal residual}
\acrodef{GPU}{graphics processor unit}
\acrodef{GS}{Gauss-Seidel}
\acrodef{GUI}{graphical user interface}
\acrodef{HPCG}{High Performance Conjugate Gradient}
\acrodef{HDL}{hardware description language}
\acrodef{HHG}{hierarchical hybrid grid}
\acrodef{HLS}{high-level synthesis}
\acrodef{HPC}{high-performance computing}
\acrodef{IACA}{Intel Architecture Code Analyzer}
\acrodef{IP}{intellectual property}
\acrodef{ISA}{instruction set architecture}
\acrodef{ITAC}{Intel Trace Analyzer and Collector}
\acrodef{IR}{intermediate representation}
\acrodef{JIT}{just-in-time}
\acrodef{KPM}{Kernel Polynomial Method}
\acrodef{KM}{Kuramoto Model}
\acrodef{LC}{Layer Condition}
\acrodef{LFA}{local Fourier analysis}
\acrodef{LBM}{Lattice Boltzmann}
\acrodef{LLC}{last-level cache}
\acrodef{LoC}{lines of code}
\acrodef{LZR}{Leibniz Supercomputing Centre}
\acrodef{MB}{memory-bound}
\acrodef{MC}{memory controller}
\acrodef{MPI}{Message Passing Interface}
\acrodef{NDG}{nodal discontinuous Galerkin}
\acrodef{NDGTD}{nodal discontinuous Galerkin time domain}
\acrodef{NIC}{network interface controller}
\acrodef{OMP}{OpenMP}
\acrodef{NT}{non-temporal}
\acrodef{NUMA}{non-uniform memory access}
\acrodef{OS}{operating system}
\acrodef{OSACA}{Open-Source Architecture Code Analyzer}
\acrodef{P2P}{point-to-point}
\acrodef{PDE}{partial differential equation}
\acrodef{RAPL}{running average power limit}
\acrodef{PGAS}{partitioned global address space}
\acrodef{PPnR}{post place and route}
\acrodef{PPS}{processes per socket}
\acrodef{QDR}{quad data rate}
\acrodef{RAM}{random access memory}\acuse{RAM}
\acrodef{RBGS}{red-black Gauss-Seidel}
\acrodef{RDMA}{remote direct memory access}
\acrodef{RHS}{right-hand side}
\acrodef{RRZE}{Regional Computer Center Erlangen} 
\acrodef{RTL}{register transfer level}
\acrodef{SHM}{shared memory}
\acrodef{SPIR}{standard portable intermediate representation}
\acrodef{SPL}{software product lines}
\acrodef{SpMV}{sparse matrix-vector multiplication}
\acrodef{SpMVM}{sparse matrix-vector multiplication}
\acrodef{SIMD}{single instruction, multiple data}
\acrodef{SMP}{symmetric multiprocessing}
\acrodef{SMT}{simultaneous multithreading}
\acrodef{SP}{single precision}
\acrodef{SoA}{Structure of Arrays}
\acrodef{SSE}{streaming SIMD extensions}
\acrodef{STL}{Standard Template Library}
\acrodef{TDP}{thermal design power}
\acrodef{TLB}{translation lookaside buffer}
\acrodef{TPDL}{target platform description language}
\acrodef{UFS}{Uncore frequency scaling}
\acrodef{WF}{wavefront}
\acrodef{XML}{eXtensible Markup Language}

\newcommand{\CPP}{C\nolinebreak[4]\hspace{-.05em}\raisebox{.23ex}{\relsize{-1}{++}}}

\newif\iftitle
\titletrue

\newcommand{\bq}{\begin{equation}}
\newcommand{\eq}{\end{equation}}
\newcommand{\bytes}{\mbox{B}}
\newcommand{\byte}{\mbox{byte}}
\newcommand{\second}{\mbox{s}}

\newcommand{\seconds}{\mbox{s}}
\newcommand{\flop}{\mbox{flop}}
\newcommand{\flops}{\mbox{flops}}

\newcommand{\bit}{\mbox{bit}}
\newcommand{\bits}{\mbox{bits}}

\newcommand{\GBS}{\mbox{GB/\second}}

\newcommand{\GFS}{\mbox{G\flop/\second}}

\newcommand{\GHZ}{\mbox{GHz}}

\newcommand{\BF}{\mbox{\byte/\flop}}

\newcommand{\GB}{\mbox{GB}}

\newcommand{\MB}{\mbox{MB}}

\SaveVerb{HHQ-large}.HHQ-large.
\SaveVerb{HHQ-small}.HHQ-small.
\SaveVerb{Emmy}.Emmy.
\SaveVerb{Meggie}.Meggie.
\SaveVerb{HazelHen}.Hazel Hen.
\SaveVerb{SuperMUCNG}.SuperMUC-NG.
\SaveVerb{Hawk}.Hawk.
\SaveVerb{MPIcommsize}.MPI_Comm_size.
\SaveVerb{MPIwaitall}.MPI_Waitall.
\SaveVerb{MPIwait}.MPI_Wait.
\SaveVerb{MPIisend}.MPI_Isend.
\SaveVerb{MPIsend}.MPI_Send.
\SaveVerb{MPIirecv}.MPI_Irecv.
\SaveVerb{MPIrecv}.MPI_Recv.
\SaveVerb{MPIsendrecv}.MPI_Sendrecv.
\SaveVerb{MPIbarrier}.MPI_Barrier.
\SaveVerb{MPIallreduce}.MPI_Allreduce.
\SaveVerb{MPIreduce}.MPI_Reduce.
\SaveVerb{MPIalltoall}.MPI_Alltoall.
\SaveVerb{MPIallgather}.MPI_Allgather.
\SaveVerb{MPIgather}.MPI_Gather.
\SaveVerb{MPIscatter}.MPI_Scatter.
\SaveVerb{MPIbcast}.MPI_Bcast.
\SaveVerb{MPI}.MPI.
\SaveVerb{OpenMP}.OpenMP.
\SaveVerb{MPIOpenMP}.MPI+OpenMP.
\SaveVerb{NUMA}.NUMA.
\SaveVerb{STREAM}.STREAM.
\SaveVerb{Cosine}.Cosine.
\SaveVerb{DIV}.DIV.
\SaveVerb{vdivpd}.vdivpd.
\SaveVerb{SpMV}.SpMV.
\SaveVerb{static}.static.
\SaveVerb{LIKWID}.LIKWID.
\SaveVerb{likwidperfctr}.featureA.
\SaveVerb{Chrono}.Chrono.
\SaveVerb{ccNUMA}.ccNUMA.
\SaveVerb{BLAS1}.BLAS-1.
\SaveVerb{BLAS2}.BLAS-2.

\newcommand{\GHcomm}[1]{{\color{red}#1\color{black}}}

\makeatletter
\renewcommand{\todo}[2][]{\@todo[caption={#2},#1]{\begin{spacing}{0.5}\fontfamily{phv}\fontseries{mc}\selectfont{#2\vspace{-1em}}\end{spacing}}}
\makeatother

\SetSymbolFont{letters}{bold}{OML}{cmm}{b}{it}
\SetSymbolFont{operators}{bold}{OT1}{cmr}{bx}{n}

\ifCLASSOPTIONcompsoc
\usepackage[nocompress]{cite}
\else
\usepackage{cite}
\fi
\ifCLASSINFOpdf
\else
\fi
\usepackage{xurl}

\newif\ifblind
\blindfalse

\begin{document}
\title{The Role of Idle Waves, Desynchronization, and Bottleneck Evasion in the Performance of Parallel Programs}

\ifblind
\author{Authors omitted for double-blind review process}
\else
%
%
\author{Ayesha Afzal, Georg Hager, and Gerhard Wellein
	\IEEEcompsocitemizethanks{\IEEEcompsocthanksitem A. Afzal, G. Hager and G. Wellein are with the Erlangen National High Performance Computing Center (NHR@FAU)\\ Friedrich-Alexander-Universit\"at Erlangen-N\"urnberg, Germany.\protect\\
	E-mail: \{ayesha.afzal, georg.hager, gerhard.wellein\}@fau.de
	\IEEEcompsocthanksitem G. Wellein is with Department of Computer Science, Friedrich-Alexander-Universit\"at Erlangen-N\"urnberg, Germany.}
	}
\fi


\IEEEtitleabstractindextext{%
  \begin{abstract}
    The performance of highly parallel applications on distributed-memory systems is influenced by many
    factors. Analytic performance modeling techniques aim to provide insight into performance limitations
    and are often the starting point of optimization efforts. However, coupling
    analytic models across the system hierarchy (socket, node, network) fails to encompass
    the intricate interplay between the program code and the hardware, especially when
    execution and communication bottlenecks are involved. In this paper we investigate
    the effect of \emph{bottleneck evasion} 
    and how it can lead to automatic overlap of communication overhead with computation.
    Bottleneck evasion leads to a gradual loss of the initial 
    bulk-synchronous behavior of a parallel code so that its processes become desynchronized. This
    occurs most prominently in memory-bound programs, which is why we choose memory-bound
    benchmark and application codes, specifically an MPI-augmented STREAM Triad, sparse matrix-vector
    multiplication, and a collective-avoiding Chebyshev filter diagonalization code
    to demonstrate the consequences of desynchronization
    on two different supercomputing platforms.
    We investigate the role of idle waves as possible triggers for desynchronization
    and show the impact of automatic asynchronous communication for a spectrum of code properties and parameters,
      such as saturation point, matrix structures, domain decomposition, and communication concurrency.
    Our findings reveal how eliminating synchronization points (such as collective communication
    or barriers) precipitates performance improvements that go beyond what can be
    expected by simply subtracting the overhead of the collective from the overall
    runtime.
  \end{abstract}
	
}


\maketitle	
\IEEEdisplaynontitleabstractindextext
\IEEEpeerreviewmaketitle
\pgfkeys{/pgf/number format/.cd,1000 sep={\,}}

\IEEEraisesectionheading{\section{Introduction}\label{sec:introduction}}
\IEEEPARstart{W}{hite-box} (i.e., first-principles) performance modeling of distributed-memory applications on multicore clusters is notoriously imprecise due to a wide spectrum of random disturbances
whose performance impact is multi-faceted.
Possible sources are system noise, variations in network performance, application load imbalance, and contention on shared resources.
Among the latter, the memory interface of a processor on a ccNUMA domain and the network interface of a compute node are only the most prominent ones.
The typical ``lock-step'' pattern of many parallel programs in computational science, where computation phases alternate with communication in a regular way, can be destroyed by these effects. We call this process \emph{desynchronization}. 
It arises even if the application is completely balanced and all computation and communication phases take the same amount of time on all processors, breaking the inherent translational symmetry of the underlying software and hardware. As a consequence, simply adding modeled computation and communication times does not yield reliable runtime predictions. 
There have been numerous efforts to assess, categorize, and reduce disturbances.
In contrast, there has been relatively little work on studying \emph{disturbance propagation} across the processes of MPI-parallel programs, which goes under the name of \emph{idle waves}.
The propagation speed of idle waves and the dynamics of their interaction with the system (e.g., with natural system noise or with each other) can be modeled accurately in some cases. In the presence of bottlenecks, desynchronization can be initiated by the presence of idle waves and lead to a stable, persistent desynchronized state, which we call \emph{computational wavefront}.
In stark contrast to the general wisdom that perfect synchronization is always desirable, the computational wavefront state can lead to a better utilization of the bottlenecked resource by automatic overlap of communication overhead with useful work. This mechanism depends on the presence of a bottleneck among processes and is influenced by the ``strength'' of the bottleneck, i.e., how many processes are needed to saturate it. Canonical examples are the memory bandwidth on a ccNUMA domain and the network connection of a compute node.
Still,
a full quantitative understanding of the performance consequences of \emph{desynchronization}, \emph{idle wave propagation}, and \emph{computational wavefronts} is still lacking, as is their impact on real-world applications.

In this paper we address the latter by investigating distributed-memory bulk-synchronous parallel benchmarks that perform computation and communication in a ``lock-step'' pattern. They interact with either contended or scalable resources available across the allocated set of compute nodes.
The definition of contention is as follows: We assume that the $N$ MPI processes access multiple shared resources, such as memory bandwidth, shared cache bandwidth, node network injection bandwidth, or even the full-system bisection bandwidth. Each resource $i$ applies to a group of $N_i$ MPI processes, which leads to $N/N_i$ \emph{contention groups}. For example, on a 20-core CPU with a single ccNUMA domain per package we have $N_\mathrm{memBW}=20$ if one MPI process is running per core. A program is \emph{resource scalable} if there are no contention groups. This can happen because there are no contended resources or because the execution mode of the program does not expose them. In the example above, if one MPI process with 20 OpenMP threads is running per ccNUMA domain, there is no memory bandwidth contention group among processes.\footnote{This does not mean that there is no memory bandwidth contention; it just plays no role for desynchronization phenomena.} 

	\begin{table}[t]
		\centering
		\caption{Programs and the investigated parameter space.} 
		\label{tab:app}
		\begin{adjustbox}{width=0.49\textwidth}
			\begin{threeparttable}
				\setlength\extrarowheight{-0.7pt}
\setlength\tabcolsep{2pt}
\arrayrulecolor{blue}
\begin{tabular}[fragile]{yz}
	\toprule
	\rowcolor[gray]{0.9}
	Parallel codes  & Parameter space for analysis\\
	\midrule
	STREAM Triad  & provoked and spontaneous disturbances  \\    
	SpMVM& matrix topology, communication concurrency    \\
	ChebFD   & communication scheme,  domain decomposition \\
	\bottomrule
\end{tabular}

			\end{threeparttable}
		\end{adjustbox}
	\end{table}
	
Table~\ref{tab:app} shows an overview of the microbenchmarks and the two mini-apps under investigation and which part of the parameter space we investigate for each. The selection aims to provide a spectrum of program properties and how idle waves and desynchronization impact their performance. 
The strongly memory-bound STREAM Triad code was augmented from its original~\cite{mccalpin1995memory} by adding next-neighbor communication in order to construct the cleanest possible setup that can show the desired effects.
The SpMVM benchmark issues back-to-back sparse matrix-vector multiplications and thus represents many sparse iterative algorithms. It is also memory bound on the node level but adds additional complexity via problem-dependent communication patterns.
Finally, in case of Chebyshev Filter Diagonalization (ChebFD)~\cite{pieper2016high}, all collectives can be eliminated from the algorithm without limiting its functionality, which makes it a representative of the growing field of communication-avoiding algorithms~\cite{carson2015communication,GHYSELS2014224}. The choice of a blocking factor also allows to fine-tune its computational intensity. 
In Sect.~\ref{sec:evaluation}, each program will be described in more detail. We
restrict ourselves to pure MPI applications; the basic phenomenology
of hybrid MPI+OpenMP codes in terms of desynchronization and have been
addressed in~\cite{AfzalHW20}.

\subsubsection*{Relevance and generalization of desynchronization}

Desynchronization is a very common phenomenon in parallel programs.
As shown in previous work~\cite{AfzalHW19,AfzalHW20}, the presence of a
resource bottleneck is decisive for the initial lock-step mode
to become unstable even under natural system noise without any
explicit disturbance from the outside.
We call this instability \emph{bottleneck evasion.}
In absence of globally
synchronizing operations,
an initially small deviation from lock-step is allowed to evolve
into an eventually stable pattern. This is especially easy to observe
in implementations of algorithms such as
synchronization-free polynomial filters\cite{Kreutzer:2015,kreutzer2018chebyshev}
or communication-avoiding Krylov subspace methods
\cite{GHYSELS2014224,carson2015communication}. However, even in the presence
of collectives it is possible for idle waves and the desynchronized
pattern to survive, depending on the particular implementation
of the collective~\cite{AfzalHW2021}. In the desynchronized state,
different loop kernels with different behavior towards chip-level
bottlenecks can execute concurrently on different cores, or execution
of code can overlap with communication-induced waiting time. The
performance impact of desynchronization for the whole program mainly
depends on the saturation characteristics of the relevant hardware bottlenecks and where the system eventually settles in terms of how many cores execute which code at any point in time.
If back-to-back program phases
with different behavior towards a common bottleneck overlap in time,
their characteristics also govern the further evolution of the
desynchronized state~\cite{AfzalHWcpe22}.
This phenomenon is not
restricted to standard multicore architectures; it is also common
in task-parallel programs and in GPUs where threads execute different
kernels in parallel~\cite{zhao2020hsm}.

\subsubsection*{Idle waves and computational wavefronts}

In prior work we also presented a validated analytic model for the {idle wave} propagation velocity~\cite{AfzalHW19,AfzalHW2021}, which shows the influence of execution and communication properties of the application, with a special emphasis on sparse communication patterns.
The propagation speed is the number of processes the delay travels per time unit, and it is measured in ranks per second.
We pointed out how idle waves decay under system noise and due to topological differences in communication characteristics among parts of the system.

In the presence of a memory bandwidth bottleneck, the propagation of {idle waves} is superimposed by the bandwidth limitations, which cause a decay of the wave over time~\cite{AfzalHW20,AfzalEuroMPI19Poster}.
In this scenario, a deliberate injection of an idle period can initiate an idle wave which, after its eventual disappearance, leaves an ``echo'' in the form of a desynchronized computational wavefront.
Hence, adding some delay on one of the processes can accelerate the transition to the desynchronized state, possibly leading to better overall time to solution.
This constitutes the important connection between idle waves and desynchronization.

%
     
\subsection{Related Work}
	
\subsubsection{Noise} 

Extensive research~\cite{petrini2003case,jones2003improving,terry2004improving,gioiosa2004analysis,tsafrir2005system,beckman2006influence,ferreira2008characterizing,morari2011quantitative,Weisbach:2018} has been conducted for almost two decades to characterize noise, identify sources of noise outside of the control of the application, and pinpoint its influence on collective operations. Explicit techniques for asynchronous communication, mitigation of noise, MPI process placement, dynamic load balancing, synchronization of OS influence, lightweight OS kernels, etc., have been explored.
In contrast, the present paper investigates the favorable consequences of noise as an enabling factor for desynchronization and -- in case of parallel programs with bottleneck(s) among processes -- automatic partial or full overlap of communication and computation.
Hoefler et al.~\cite{hoefler2010characterizing} used a simulator based on the LogGOPS communication model to investigate both \ac{P2P} and collective operations to study the influence of system noise on large-scale applications.
They found that application scalability is mostly determined by the noise pattern and not the noise intensity.
However, their data was not taken on a real cluster and it is neither aware of node-level bottlenecks nor does it take the system topology and different kinds of delay propagation into account.
In the specific context of idle wave decay, Afzal et al.~\cite{AfzalHW2021} found that the noise intensity is the main influence factor rather than its detailed statistics.
	
\subsubsection{Parallel computing dynamics}

There is very little research on idle wave propagation and spontaneous pattern formation in parallel code, especially in the context of memory-bound programs.
Hence, none of the existing prior work addressed spontaneous pattern formation and desynchronization.
Markidis et al.~\cite{markidis2015idle} used a LogGOPS simulator~\cite{hoefler2010characterizing} for a phenomenological study of idle wave propagation. They concluded that isolated idle periods propagate among MPI processes as nondispersive, damped linear waves.
However, their simulator is neither aware of communication topology, concurrency, and mode (rendezvous vs.\ eager) nor does it consider the socket-level character of the code and the quantitative investigation of the connection between damping and noise.
Their speculation that idle waves are described by a linear cannot be upheld~\cite{AfzalHW19}. 
Gamell et al.~\cite{Gamell:2015} observed the emergence of idle periods in the context of failure recovery and failure masking of stencil codes. Boheme et al.~\cite{Boehme:2016} presented a tool-based approach to attribute propagating wait states in MPI programs to their original sources, helping to identify and correct the root issues. Kolakowska et al.~\cite{kolakowska2004desynchronization} carried out a study of the virtual time horizon in conservative parallel discrete-event simulations (PDES). However, in all studies, the global properties of such waves, like damping and velocity, and the interaction with memory-bound characteristics of the application were ignored.

Afzal et al.~\cite{AfzalHW19,AfzalHW2021,AfzalEuroMPI19Poster,AfzalHW20,AfzalHW2021,AfzalISC21Poster} were the first to investigate the dynamics of idle waves for a variety of communication patterns, (de)synchronization processes, and computational wavefront formation in parallel programs with core-bound and memory-bound code, showing that nonlinear processes dominate there.
It turned out that on real cluster systems with their complex node-level topology, MPI-parallel memory-bound programs exhibit local, non-static idles waves (variable frequency and speed) which ripple through the MPI processes. These disturbances ultimately settle down into a global steady state, the computational wavefront.
This transition time is dependent on numerous factors, such as communication volume, system size, seeds for naturally occurring one-off disturbances (``kicks''), etc.
Our work takes up from this point and extends it towards investigating the influence of such dynamics on the performance of real-world distributed-memory parallel codes. 

\subsubsection{Performance modeling and optimization}

Performance modeling is a powerful tool to investigate the properties of code in order to get insight into its bottlenecks and, consequently, identify optimization opportunities. 
In contrast to white-box (i.e., first-principles) performance models, accurate black-box approaches for describing the performance and scalability of highly parallel programs have been available for some time. These typically employ curve fitting, machine learning, and general AI methods~\cite{an2011score,Neural2019}.
A lot of related research has also been done on code optimization, focusing on new data structures, efficient algorithms, and parallelization techniques, all of which require explicit programming~\cite{hager2010introduction}.
In the present paper we investigate a specific mechanism -- automatic communication overlap -- which does not need any code changes.
Note that this does not mean that ``traditional'' means of ensuring communication overlap such as implicit or explicit asynchronous progress threads, DMA transfers, etc., become non-essential. Depending on the details, automatic overlap may not be able to hide all the communication cost or -- in general -- settle in the most favorable state in terms of time to solution. What we want to highlight here is the impact of automatic desynchronization on performance in terms of easily obtainable metrics and a well-defined experimental procedure, and to identify which properties of the code-machine interaction influence this effect.
%
			
\subsection{Contribution}

This paper tries to fathom the role of desynchronization in parallel application programs by analyzing microbenchmarks and popular proxy applications. It makes the following relevant contributions:
\begin{itemize}[\setlength\topsep{0pt}]
\item The slope of a computational wave, i.e., the degree of desynchronization among processes, is directly correlated with idle wave speed on the same system. This serves as a guiding principle to assess desynchronization phenomena in applications.
  \item An established computational wavefront can be shifted through the system (along the process direction) by deliberately injecting a delay which, however, has to be strong enough to provoke the change.
  \item Slow idle wave propagation speed caused by a small matrix bandwidth facilitates automatic overlap of communication and computation in SpMVM.
  \item In ChebFD, the particular domain decomposition and its impact on communication distances and idle wave speed is decisive for automatic communication overlap. 
  \end{itemize}
	\begin{table}[t]
		\centering
		\caption{ Key hardware and software traits of systems.} 
		\label{tab:system}
		\begin{adjustbox}{width=0.49\textwidth}
			\begin{threeparttable}
				\setlength\extrarowheight{-0.7pt}
\setlength\tabcolsep{2pt}
\arrayrulecolor{blue}
\Huge
\begin{tabular}[fragile]{c>{~}lwx}
	\toprule
	\rowcolor[gray]{0.9}
	\cellcolor[gray]{0.9}&Systems  & Meggie (M)   &  SuperMUCNG (S)  \\
	\midrule
	\cellcolor[gray]{0.9}&Processor  & Intel Xeon Broadwell EP   & Intel Xeon Skylake SP   \\    
	\cellcolor[gray]{0.9}&Processor Model      & E5-2630 v4   & Platinum 8174     \\
	\cellcolor[gray]{0.9}&Base clock speed &\SI{2.2}{\giga \hertz}    &  \SI{3.10}{\giga \hertz} (\SI{2.3}{\giga \hertz} used\mbox{$^\ast$})   \\
	\cellcolor[gray]{0.9}&Physical cores per node    & 20   & 48          \\
	\cellcolor[gray]{0.9}&Numa domains per node  &   2  & 2     \\
	\cellcolor[gray]{0.9}&Last-level cache (LLC) size & \SI{25}{\mega \byte} (L3)  & \SI{33}{\mega \byte} (L3) + \SI{24}{\mega \byte} (L2)     \\
	\cellcolor[gray]{0.9}&Memory per node (type)& \SI{64}{\giga \byte} (DDR4)  & \SI{96}{\giga \byte} (DDR4)   \\
	\multirow{-8}{*}{\rotatebox{90}{\cellcolor[gray]{0.9} Micro-architecture}}&Theor. memory bandwidth & \SI{68.3}{\giga \byte / \second}&  \SI{128}{\giga \byte / \second}\\
	
	\midrule
	\cellcolor[gray]{0.9}&Node interconnect    & Omni-Path  & Omni-Path      \\
	\cellcolor[gray]{0.9}&Interconnect topology & Fat-tree & Fat-tree \\
	\multirow{-3}{*}{\rotatebox{90}{\cellcolor[gray]{0.9} Network}}&Raw bandwidth p. lnk n. dir &\SI{100}{\giga \bit \per \second}  &    \SI{100}{\giga \bit \per \second}   \\
	
	\midrule
	\cellcolor[gray]{0.9}&Compiler    & Intel \CPP{} v2019.5.281\mbox{$^\ddag$}     & Intel \CPP{} v2019.4.243     \\
	\cellcolor[gray]{0.9}&Optimization flags & -O3 -xHost  & -O3 -qopt-zmm-usage=high \\
	\cellcolor[gray]{0.9}&SIMD & -xAVX &-xCORE-AVX512\\
	\cellcolor[gray]{0.9}&Message passing library & Intel \verb.MPI. v2019u5     & Intel \verb.MPI. v2019u4     \\
	\multirow{-5}{*}{\rotatebox{90}{\cellcolor[gray]{0.9} Software}}&Operating system    & CentOS Linux v7.7.1908  &  SUSE Linux ENT. Server 12 SP3    \\
	
	\midrule
	\cellcolor[gray]{0.9}&\verb.ITAC.    & v2019u5      & v2019   \\
	\multirow{-2}{*}{\rotatebox{90}{\cellcolor[gray]{0.9} Tools}}&\verb.LIKWID.   & 5.0.1         & 5.0.1     \\
	\bottomrule
\end{tabular}

				\begin{tablenotes}[flushleft]
					\item \mbox{$^\ast$} A power cap is applied on
					\UseVerb{SuperMUCNG}, i.e., the CPUs run by
					default on a lower than maximum clock speed (\SI{2.3}{\giga
						\hertz} instead of \SI{3.10}{\giga \hertz}).
					\item \mbox{$^\ddag$} \texttt{oneapi/2021.1.1} is used, whenever specified specifically.
				\end{tablenotes}
			\end{threeparttable}
		\end{adjustbox}
	\end{table}
	This paper is organized as follows:
	We first provide details about
	our experimental environment and methodology in Sect.~\ref{sec:environment}.
	In Sect.~\ref{sec:evaluation} we discuss the performance phenomenology and implications
        of desynchronization on a stream-like microbenchmark, sparse matrix-vector
        multiplication (SpMVM), and a Chebyshev filter diagonalization application
        (ChebFD). In Sect.~\ref{sec:conclusion} we summarize the paper and
        give an outlook to future work.

\begin{figure}[t]
	\centering
	\includegraphics[scale=0.75]{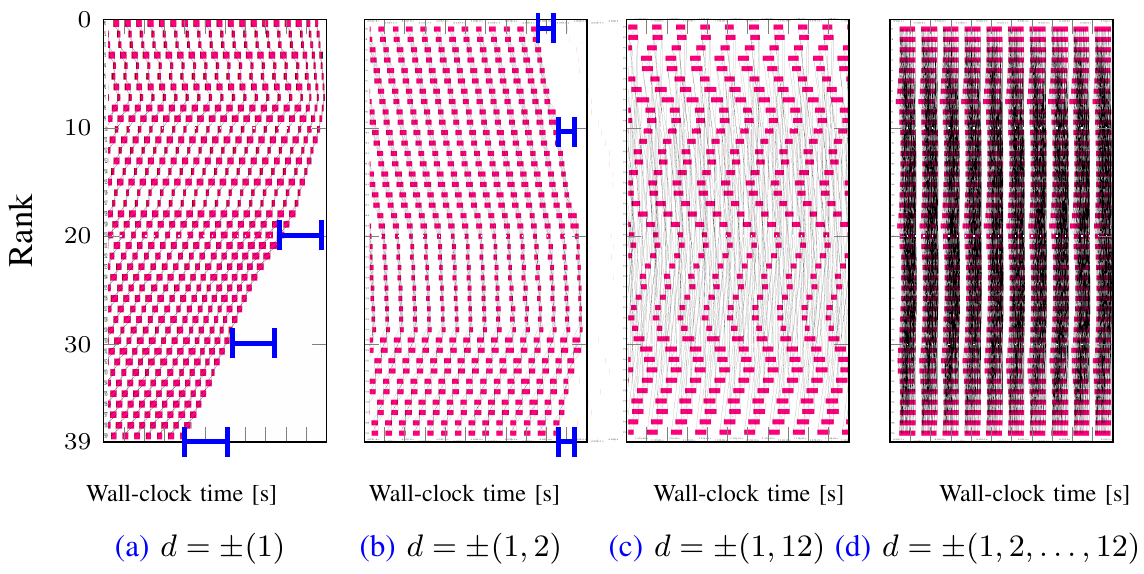}
	\caption{Snippets of MPI trace timelines of a fully developed
		computational wavefront state in the MPI-augmented
		STREAM Triad with different communication topologies
		(\textcolor{blue}{(a)--(d)}), open boundary conditions, and $5\times
		10^4$ iterations on 40 cores (four ccNUMA domains)
		of the \protect\UseVerb{Meggie} cluster. Only the
		waiting time spent in the MPI library (red) and the
		message transfers (black lines) are shown. 
	} 
	\label{fig:Wavefronts}  
\end{figure}
\section{Test bed and Experimental Setup} \label{sec:environment}

\subsection{HPC platforms and architectures}
To ensure the broad applicability of our results, we conduct most
experiments on two different clusters:
\begin{enumerate}[\setlength\topsep{0pt}]
\item SuperMUC-NG\footnote{\url{https://doku.lrz.de/display/PUBLIC/SuperMUC-NG}},
  an Omni-Path cluster with two Intel Xeon ``Skylake SP'' CPUs per node and
  24 cores per CPU (hyper-threading enabled)
\item Meggie\footnote{\ifblind{URL redacted for double-blind review}\else\url{https://anleitungen.rrze.fau.de/hpc/meggie-cluster}\fi},
  an Omni-Path cluster with two Intel Xeon
  ``Broadwell'' CPUs per node and 10 cores per CPU (hyper-threading disabled)
\end{enumerate}
Both systems have significant differences in the numbers of cores per ccNUMA domain and their memory bandwidth. Although hyper-threading is active on SuperMUC-NG, we ignore it in this work.
Further details of the hardware and software environments
can be found in \Cref{tab:system}.
	
	
\subsection{Parameters, notations, and methodology}
Runtime traces were visualized using the Intel Trace Analyzer and Collector (ITAC) tool.
Process-core affinity was enforced using the 
\texttt{I\_MPI\_PIN\_PROCESSOR\_LIST} environment variable.
The clock frequency was always fixed 
(2.2\,\GHZ\ base clock speed on Meggie and 2.3\,\GHZ\ in case of \UseVerb{SuperMUCNG} because of
the power capping mechanism).
Unless otherwise noted, to enable overlap via hiding communication in parallel with computation supported by the MPI implementation, we set the \texttt{I\_MPI\_ASYNC\_PROGRESS} environment variable to one  and use the Intel MPI multi-threaded optimized  \texttt{release\_mt}\footnote{\url{https://software.intel.com/content/www/us/en/develop/documentation/mpi-developer-reference-linux/top/environment-variable-reference/environment-variables-for-asynchronous-progress-control.html}} library, which supports asynchronous progress threads.
Working sets for memory-bound cases were chosen large enough to not fit into the available cache, i.e., at least 10$\times$ the size of all \ac{LLC}, which is the L3 caches (non-inclusive victim L3 caches) in the Broadwell (Skylake) processors of Meggie (SuperMUC-NG).
All floating-point computations are done in double precision.
Individual kernel executions were repeated at least
15 times to even out variations in runtime. We report statistical fluctuations if they were significant.

\subsection{Desynchronization speedup metric} \label{sec:evalMatrix}

We use performance measurements to quantify the speedup caused by
desynchronization via bottleneck evasion.
Experimentally, we determine the quantity 
\bq\label{eq:Dfactor_exp}
P_D = \frac{P_\mathrm{barrier\_free}-P_\mathrm{barrier}}{|P_\mathrm{barrier}|}~, 
\eq
which measures the speedup of the program when eliminating an artificial barrier synchronization. 
To make this comparison useful, the actual barrier overhead (as measured by a microbenchmark)
is subtracted from the overall runtime before calculating $P_\mathrm{barrier}$.
A higher $P_D$ factor indicates a more effective communication overlap and better scalability.
			
\section{Evaluation and Implications} \label{sec:evaluation}

This section describes analysis results for the selected microbenchmarks and proxy applications with a focus on idle waves and the automatic overlap communication and computation via computational wavefronts. The goal is not to provide a comprehensive analysis of each application but an assessment of desynchronization impact. Great care has been taken to separate the speedup observed by desynchronization from other desirable effects, such as the removal of synchronizing collectives or the reduction of communication volume.

	
	\begin{figure}[t]
		\centering
		\includegraphics[scale=1]{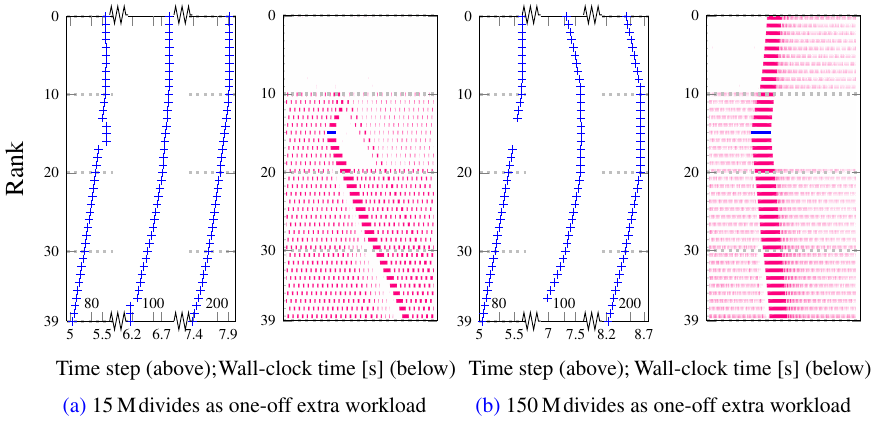}
		\caption{Stability of computational wavefronts
			against (a) small and (b) large disturbances.  The slope
			of the wavefront is the same (\SI{60}{ranks
				\per \second}) before the extra workload is
			injected (79th iteration) and after the idle wave
			dies out (200th iteration).
			{\textcolor{blue}{(a)}} The ``lagging'' ccNUMA domain
			(first socket) remains the same for a small
			disturbance.  {\textcolor{blue}{(b)}} The ``lagging''
			ccNUMA domain shifts to the second
			socket, which is where the large disturbance was injected.
		}
		\label{fig:WavefrontStability}
	\end{figure}
	
	\begin{table*}[t!]
		\centering
		\caption{Structure of the MPI-parallel SpMVM implementation. Split-wait and non-split are implementation alternatives.}
		\label{tab:SPMValgo}
		\begin{threeparttable}
			\includegraphics{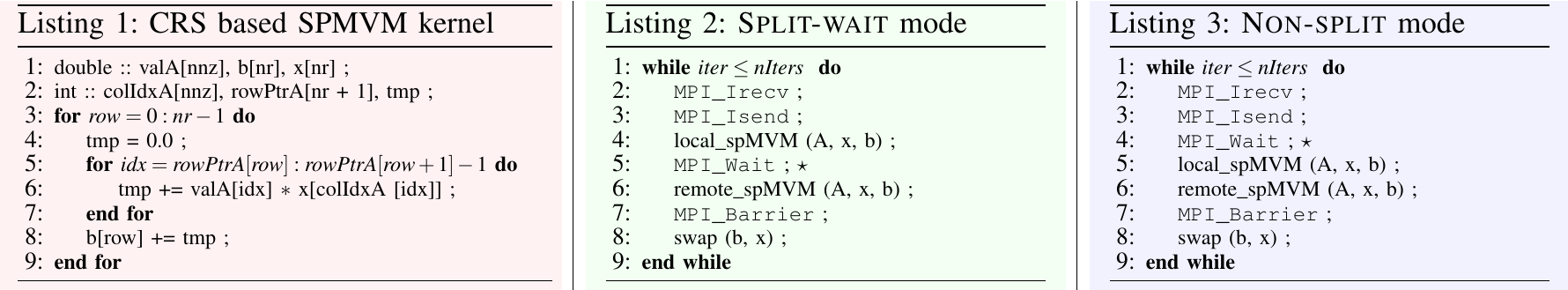}
			\begin{tablenotes}[flushleft]
				\footnotesize
				\item \mbox{$^\star$}
				Two \UseVerb{MPIwait} routines wait for both MPI receive and send requests to complete.
			\end{tablenotes}
		\end{threeparttable}
	\end{table*}
	
	
\subsection{MPI-augmented STREAM Triad} \label{sec:TRIAD}

We start with the simple case of the pure-MPI version of the McCalpin STREAM Triad~\cite{mccalpin1995memory} loop (\verb.A(:)=B(:)+s*C(:).). It allows
a straightforward application of the Roof{}line model to predict the memory-bound
parallel performance limit on a ccNUMA domain as $P=b_\mathrm S/B_\mathrm C$,
where $b_\mathrm S$ is the domain memory bandwidth and $B_\mathrm c=12\,\BF$ is the
code balance (assuming that streaming stores are used, i.e., write-allocate
transfers do not apply). A constant overall working set of \SI{2.4}{\giga\bytes}~($10^8$ elements) is distributed evenly among the MPI processes. Communication is added after each sweep of the loop
to mimic a real MPI-parallel program. The communication topology can be varied, but non-blocking point-to-point calls (\texttt{MPI\_Isend}/\texttt{MPI\_Irecv}) together with a final \texttt{MPI\_Waitall} are used in all cases.
The implementation uses open boundary conditions (i.e., process $0$ ($n-1$) only communicates with processes $1$ and above ($n-2$ and below) and bidirectional direct-neighbor communication (i.e., each MPI process $i$ sends and receives \SI{16384}{\bytes} to and from $i\pm1$ after each STREAM phase).

\subsubsection{Shape of computational wavefronts}

%
%
%
Figure~\ref{fig:Wavefronts} shows MPI traces of the MPI-augmented STREAM Triad on 40 processes (two nodes of Meggie) with different communication topologies.  
In these figures, code execution is white and MPI waiting time is red. With pure next-neighbor communication as in (a), the number of concurrently active (i.e., code-executing) processes per ccNUMA domain settles in the vicinity of the performance saturation point\footnote{A saturation point is the minimum number of processes required to achieve the maximum memory bandwidth on the ccNUMA domain.}, as was already pointed out in~\cite{AfzalHW20}. 
In (b) one can observe the difference to next-pair communication ($d=\pm(1,2)$): The developed desynchronized state (the computational wavefront) is about $3\times$ steeper, i.e., it has a smaller amplitude. This correlates with the higher idle wave velocity for longer-distance communication scenarios~\cite{AfzalHW2021}. This restricts the communication-computation overlap, and the number of concurrently active processes per ccNUMA domain is higher than what is needed for saturation.
With a mixed short-/long-range communication topology as in (c),
computational wavefront structures on different scales overlap, i.e., two periodicities can be observed which emerge from the long- and short-distance communication, respectively. Compact long-distance communication as in (d) with $d=\pm(1,\ldots,12)$ causes high-velocity idle waves~\cite{AfzalHW2021},
which leads to steep  computational wavefronts.
Starting at a communication distance of at least $d=\pm(1,\ldots,8)$, the processes keep in lockstep for a system size of $40$ ranks as in the example. This is due to the comparatively small system size; idle waves are so fast that they leave the system in a single compute-communicate cycle.


\subsubsection{Wavefront stability}		

The question arises how stable a developed computational wavefront
is against disturbances like system noise or single one-off delay
injections. After all, the translational symmetry of the system should
not favor a particular position of the ``lagger,'' i.e., the slowest process. 
In all our measurements, natural system noise was never able
to alter the shape or position of computational wavefronts. However, long
one-off delay injections can. In  Fig.~\ref{fig:WavefrontStability} we show
the results of an experiment on the \protect\UseVerb{Meggie} system, where a fully developed computational wavefront
state is disturbed by an injection on a ccNUMA domain different from
the one where the lagger initially resides. To spark an idle wave, a series of floating-point divide operations are performed by rank $15$ at time step $80$ as illustrated by blue bars in the second and forth graph of Fig.~\ref{fig:WavefrontStability}. For one-off disturbances, we use a core-bound workload which does not impose an additional strain on the memory interface. The overall impact is independent of the nature of the disturbance though.
As a consequence, after the
ensuing idle wave has run out, the lagger
shifts to the domain where the injection took place \emph{if} the injection
is strong enough. There is currently no first-principles understanding
about what ``strong enough'' means;
experimentally, we observe that the idle wave must at least
  be able to travel
  (despite the inevitable damping) far enough as to intrude the
slowest socket.
	\begin{table*}[t!]
		\centering
		\caption{Measured walltime minimum, maximum, and median
			for execution (rows 1--3)
			and communication (rows 4--6) of one MPI-only SpMVM
			with the \protect\UseVerb{HHQ-large} matrix on
			SuperMUC-NG, using strong scaling from 96 processes (two nodes)
			up to 1296 processes (27 nodes) using barriers
			between successive SpMVMs. Row 7 shows the
			mean per-process message sizes (transmitted via rendezvous protocol at small processes count till eager limit), and the last row
			denotes the median of the communication-to-execution time ratio
			(CER). Color coding
			is used as a guide to the eye
			(white to pink scale).}
		\label{tab:SPMVECR}
		\begin{adjustbox}{width=\textwidth}
			\begin{threeparttable}
				\centering
\huge
\pgfplotstabletypeset[
/pgfplots/colormap/cool, 
every column/.style=,
color cells={min=0,max=6,textcolor=black},
/pgf/number format/fixed,
/pgf/number format/precision=2,
col sep=comma,
@content options for rows={0}{,color cells={min=1.35,max=31.58}},
@content options for rows={1}{,color cells={min=4.76,max=64.92}},
@content options for rows={2}{,color cells={min=3.03,max=53.5}},
@content options for rows={3}{,color cells={min=2.45,max=20.7}},
@content options for rows={4}{,color cells={min=10.16,max=38.73}},
@content options for rows={5}{,color cells={min=6.13,max=29.48}},
@content options for rows={6}{,color cells={min=105,max=2390}},
@content options for rows={7}{,color cells={min=0.33,max=2.077}},
]{
	{Phase vs. Rank-order},96-pe,144-pe,240-pe,480-pe,720-pe,960-pe,1296-pe,96-ep,144-ep,240-ep,480-ep,720-ep,960-ep,1296-ep
	Exec min [\si{\milli \seconds}],31.581,18.0414,9.672,4.283,3.13122,2.438,1.824,26.087,19.9199,9.72065,3.918,2.71905,1.98816,1.345
	Exec max [\si{\milli \seconds}],64.924,45.7335,27.5488,13.492,10.5258,8.254,6.236,53.3554,36.2189,22.131,11.49,7.98069,6.58592,4.76
	Exec median [\si{\milli \seconds}],53.499,35.7364,18.5135,9.051,6.530295,5.057,3.846,48.4737,30.8658,17.5683,8.2305,5.69589,4.54202,3.034
	Comm min [\si{\milli \seconds}],20.7,15.378,6.719,5.620,3.565,4.034,2.513,9.275,8.499,3.725,2.476,2.038,1.177,2.454
	Comm max [\si{\milli \seconds}],38.734,29.486,20.995,18.464,14.834,12.973,11.320,19.358,18.780,17.710,15.358,11.588,8.532,10.162
	Comm median [\si{\milli \seconds}],29.482,24.562,16.170,14.745,11.283,9.781,7.989,16.165,15.012,14.408,12.165,8.957,6.618,6.131
	Mean P2P msg size [\si{\kilo \bytes}],2390,1460,957,480,302,213,153,1310,848,505,260,178,137,105
	CER median ,0.551075721041515,0.687310417389552,0.873416695924596,1.629101756711966,1.727793307959288,1.934150682222662,2.077223088923557,0.33347980451255,0.486363548004588,0.820113499883313,1.478039001275743,1.572537390995964,1.457060955257793,2.020764667106131
}
			\end{threeparttable}
		\end{adjustbox}
	\end{table*}
	\subsection{{S}parse {M}atrix-{V}ector {M}ultiplication ({SpMVM})}\label{sec:spMVM} 		
	The multiplication of a sparse matrix with a dense vector
        ($\vec{y}=A\vec{x}$) is a central component in numerous
        numerical algorithms such as linear solvers and eigenvalue
        solvers. For large matrices, the performance of \ac{SpMVM} is
        memory bound on the node level due to its low computational
        intensity. 
        Distributed-memory
        parallelization requires the matrix $A$ and the vectors $x$ and $y$
        to be distributed across MPI processes. This can cause significant
        communication overhead if the pattern of nonzeros in the matrix is
        very scattered.

        An \ac{SpMVM} kernel is usually the dominant part of a larger
        algorithm (such as Conjugate-Gradient); sometimes, several
        \ac{SpMVM} kernels are executed in a back-to-back manner. Together
        with the properties described above, \ac{SpMVM} constitutes
        an interesting test bed for desynchronization phenomena. In this
        section we investigate such a sequence of MPI-parallel \ac{SpMVM}s, with
        left-hand side (LHS) and right-hand side (RHS) vectors swapped after
        every step. There is no explicit or implicit synchronization among
          MPI processes.
    
	\subsubsection{Implementation}
        A compressed storage format must be chosen for the sparse matrix
        so that the \ac{SpMVM} can be carried out efficiently. On multicore
        CPUs, the standard Compressed Row Storage (CRS) format is typically
        a good choice. It allows for a compact implementation of the
        kernel that enables to exploit the relevant bottleneck (memory
        bandwidth) in many cases (see Listing $1$ of Table~\ref{tab:SPMValgo})\@.
        CRS requires one-dimensional arrays for matrix entries (\verb.valA[].),
        column indices (\verb.colIdxA[].), and row pointers (\verb.rowPtrA[].)\@.
        If the matrix entries are in double precision and the indices are
        32-bit integers, the minimum code balance for CRS-\ac{SpMVM} is
        $6\,\BF$\footnote{Per iteration, the kernel carries out $2$ \flops\ and
            causes a minimum data traffic of $8$\,\byte\ for the matrix entry
            and $4$\,\byte\ for the column index.}~\cite{gropp1999toward,Moritz2014}\@.

        In the MPI-parallel \ac{SpMVM} implementation, contiguous
        blocks of matrix rows (and corresponding LHS and RHS vectors)
        are assigned to the processes so that the number of matrix
        nonzeros per process is as balanced as possible. Each process
        can compute the part of the \ac{SpMVM} for which it already
        holds the LHS and RHS entries right away. Matrix entries
        outside of this column range require communication of the
        corresponding RHS values. Splitting the operation into
        ``local'' and ``remote'' kernels causes an additional memory
        traffic of $16/n_{nzr}$\,\byte\ per multiply-add because
          the local result vector must be updated twice in memory~\cite{gropp1999toward}. 

        Two different implementations were tested:
	\begin{enumerate}[\setlength\topsep{0pt}]
        \item \textsc{\textbf{split-wait}} mode: Communication is
          initiated with non-blocking MPI calls before the local
          \ac{SpMVM} and finalized after it. Only after the
          call to  \texttt{MPI\_Wait} can the remote \ac{SpMVM} kernel 
          be executed. This allows for overlapping communication with
          the local \ac{SpMVM} if the MPI implementation supports it;
          see Listing $2$ of Table~\ref{tab:SPMValgo}.
          
        \item \textsc{\textbf{non-split}} mode: The full {non-blocking} remote
          communication is initiated and finalized before the local
          and remote \ac{SpMVM} kernels are called. This rules out any
          communication overlap by MPI; see Listing $3$ of Table~\ref{tab:SPMValgo}.
          In this case, the two kernel calls could be fused for improved
          computational intensity, but we want to keep the properties
          of the underlying kernels unchanged for the experiments shown
          here.
        \end{enumerate}
    \begin{table}[t!]
    	\centering
    	\caption{Key specifications of symmetric sparse matrices.} 
    	\label{tab:mat}
    	\begin{adjustbox}{width=0.49\textwidth}
    		\begin{threeparttable}
    			\centering
\setlength\extrarowheight{-0.7pt}
\setlength\tabcolsep{2pt}
\arrayrulecolor{blue}
\small
\begin{tabular}[fragile]{c>{~}wxyyyz}
	\toprule
	\rowcolor[gray]{0.9}
	{Matrix-order}&{Bandwidth}&{$n_\mathrm{electrons}-n_\mathrm{sites}-n_\mathrm{phonons}$\mbox{$^\mathsection$}}&{$n_\mathrm{r}=n_\mathrm{c}$\mbox{$^\star$}}	&{$n_\mathrm{nz}$\mbox{$^\star$}}	&{$n_\mathrm{nzr}$\mbox{$^\ddag$}}	&{Size [\si{\giga\bytes}]\mbox{$^\mathparagraph$}}\\
	\midrule
	HHQ-large-\emph{pe}&high&$3 - 8 - 10$&60988928&889816368&13&10.9\\
	HHQ-large-\emph{ep}&low&$3 - 8 - 10$&60988928&889816368&13&10.9\\
	HHQ-small-\emph{pe}&high&$6 - 6 - 15$&6201600&92527872&15&1.14\\
	HHQ-small-\emph{ep}&low&$6 - 6 - 15$&6201600&92527872&15&1.14\\
	\bottomrule
\end{tabular}
    			\begin{tablenotes}[flushleft]
    				\item \mbox{$^\mathsection$}
    				The described quantum system comprises $n_\mathrm{electrons}$ electrons on
    				$n_\mathrm{sites}$ lattice sites coupled to $n_\mathrm{phonons}$ phonons.
    				\item \mbox{$^\star$}
    				$n_{r}$, $n_{c}$ and $n_{nz}$
    				are the total number of rows, columns and non-zero entries of sparse square matrix respectively.
    				\item \mbox{$^\ddag$}
    				The inner loop length of the CRS SpMVM kernel $n_{nzr} (\approx\frac{n_{nz}}{n_{r}})$
    				is the average number of non-zero entries in each row of the sparse matrix.
    				\item \mbox{$^\mathparagraph$}
    				Data set size is estimated by
    				$12n_{nz}+4n_{r}$ (eight byte per matrix entry and four byte for column indices).
    			\end{tablenotes}
    		\end{threeparttable}
    	\end{adjustbox}
    \end{table}
    \begin{figure*}[t!]
    	\centering
    	\includegraphics{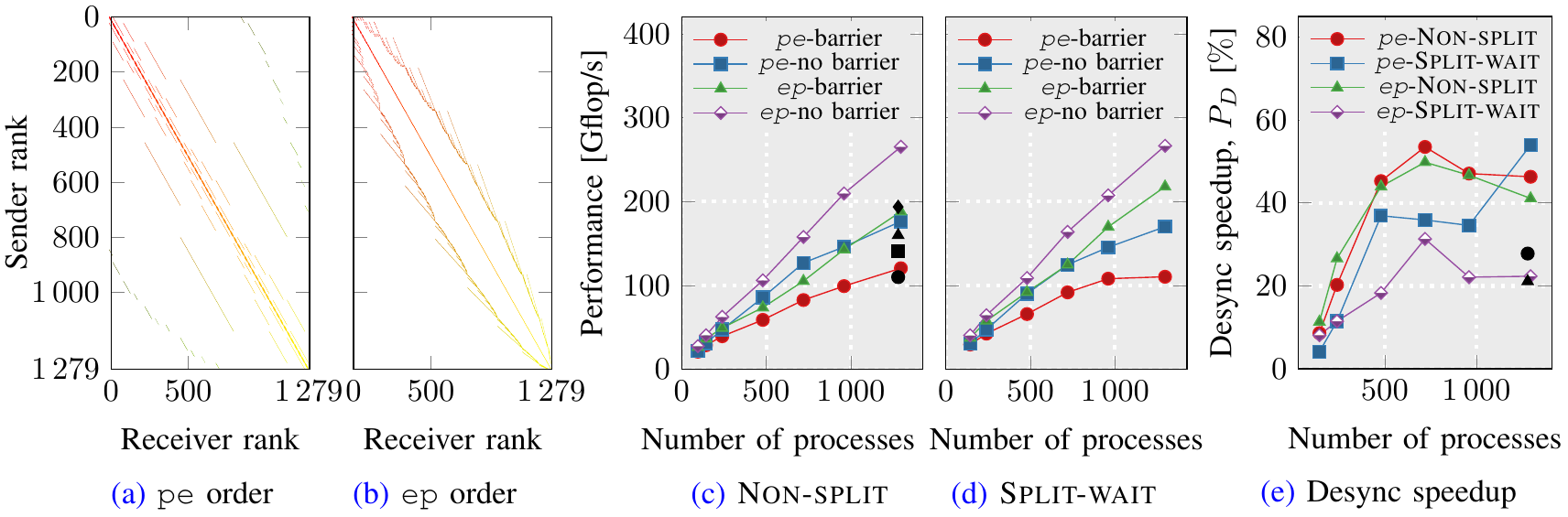}%
    	\caption{
    		{\textcolor{blue}{(a-b)}}
    		The communication topology of  \texttt{HHQ-large} matrices using periodic boundary conditions in the wider \emph{pe} order (left/right bandwidth of 41385344) and slimer \emph{ep} order (left/right bandwidth of 12907776), respectively.
    		{\textcolor{blue}{(c-d)}}
    		Strong scaling performance for $10^6$ iterations on SuperMUC-NG using \textsc{non-split} and \textsc{split-wait} algorithms, respectively. 
    		{\textcolor{blue}{(e)}}
    		Performance boost as a result of improved overlap in the absence of barriers in the \textsc{non-split} and \textsc{split-wait} implementation on the SuperMUC-NG system.
    		Markers in black at 1280 processes mark performance and the $P_D$ factor for the \textsc{non-split} algorithm on the Meggie system.
    	}
    	\label{fig:SPMV} 
    \end{figure*}
	\subsubsection{Test matrices}
        For benchmarking we use real, symmetric matrices that describe
        a strongly correlated one-dimensional electron-phonon system
        in solid state physics (Holstein-Hubbard
        Hamiltonian)~\cite{fehske2004quantum}.  The key specifications
        of the matrices are shown in \Cref{tab:mat}. Due to the
        moderate number of nonzeros per row (13 and 15, respectively),
        the minimum code balance is about 6.9\,\BF\ and 7.1\,\BF, respectively
                (assuming optimal reuse of the right-hand side vector;
          see also \cite{Moritz2014}.).
        Overall we use
        four variants that emerge from two different problem sizes
        (numbers of electrons, phonons, and lattice sites) and two
        different orderings of the degrees of freedom (phonons first
        vs.\ electrons first). The ``phonons first'' numbering
        (labeled ``pe'') produces a more scattered matrix, whereas
        with ``electrons first'' (labeled ``ep'') the nonzeros are
        closer to the diagonal (see (a) and (b) of
        Figs.~\ref{fig:SPMV} and~\ref{fig:SPMV2}).  The motivation
        behind the different problem sizes (10.9\,\GB\ and 1.135\,\GB\
        for the matrix, respectively) is that the smaller problem can
        fit into the aggregate last-level cache of the CPUs in the
        chosen clusters at a moderate node count, removing the memory
        bandwidth bottleneck at the socket level.  The matrices were
        generated using the scalable matrix collection
        (\texttt{ScaMaC}) library.\footnote{The\texttt{ScaMaC} library
          allows for scalable generation of large matrices related to
          quantum physics applications.  The open source
          implementation is available for download at
          \url{https://bitbucket.org/essex/matrixcollection/} and
          documentation of matrices can be found at
          \url{https://alvbit.bitbucket.io/scamac_docs/_matrices_page.html}.}
		
        \subsubsection{Matrix topology and communication schemes}
        The communication characteristics of distributed-memory
        \ac{SpMVM} depend strongly on the structure of the sparse
        matrix. Thus we expect the \emph{pe} versions of the
        Hamiltonians to have larger communication overhead. The
        sparsity pattern impacts the node-level performance and
        bandwidth saturation as well, however, due to the indirect
        access to the RHS vector. Table~\ref{tab:SPMVECR} shows
        execution and communication properties of one SpMVM execution
        with the large \emph{pe} and \emph{ep} matrices, respectively,
        for different numbers of MPI processes on the SuperMUC-NG
        system.
        To keep the MPI processes in lockstep, an MPI barrier
        was called before the SpMVM (the barrier time is not part of
        the reported communication time)\@.
        The data shows that the more scattered \emph{pe} matrix
        clearly causes much higher communication overhead, especially
        at lower process counts where \emph{pe} incurs more communication
        partners per rank than \emph{ep}. It can be seen 
        that the communication overhead in SpMVM
        is significant but not dominant at 96 processes. 
        The last row of the table shows the median of the
        communication-to-execution time ratio (CER), which can serve
        as a rough indicator of communication boundedness.
        The minimum, maximum, and median numbers for execution and
        communication times indicate that even in a single SpMVM without
        desynchronization there is considerable variation in both metrics
        across processes.
        
        In order to fathom the consequences of desynchronization, we
        compare the barrier version of the benchmark (i.e., a barrier
        after each SpMVM) with the barrier-free version.  Performance
        for the barrier version was calculated by subtracting the
        actual barrier time (as determined by a separate benchmark)
        from the measured walltime. Any observed speedup of the
        barrier-free version must thus be caused by automatic overlap
        of communication via desynchronization of
        processes. Figures~\ref{fig:SPMV}(c) and (d) show strong
        scaling performance for the \texttt{HHQ-large} matrices on
        SuperMUC-NG. The behavior of the split (c) and non-split (d)
        variants is similar.  Note that the best version (\emph{ep}
        without barrier) is strongly communication bound at 1296
        processes: Assuming a socket memory bandwidth of 100\,\GBS,
        the Roof{}line limit is 760\,\GFS, while the
        observed performance is only about 270\,\GFS.  The speedup $P_D$ 
        (defined in Section~\ref{sec:evalMatrix}) caused by
        bottleneck evasion via
        desynchronization is shown in Fig.~\ref{fig:SPMV}(e)\@.
        Depending on the matrix structure and the communication
        scheme, performance gains between 20\% and 55\% (out of
        a theoretical maximum of 100\%) can be observed.
        This goes with a significant improvement in scalability.

        Although the details of matrix partitioning and communication
        topology add a considerable amount of variation, the speedup
        $P_D$ shows the expected behavior along the scaling curve: It
        starts out small because the communication overhead is small
        (albeit significant), providing only minor opportunity for
        overlapping.  As the number of processes grows, this benefit
        becomes larger until at some point communication and
        computation take roughly the same amount of time.  This is
        when no further speedup can be expected. Scaling up further,
        the benefit drops because communication is dominant. One can
        also see that the non-split communication scheme (circles and
        triangles) generally shows higher speedup than the split-wait
        scheme (squares and diamonds). This is expected because
        no-split has no potential for asynchronous MPI communication
        in the lockstep case; this leaves more opportunity for overlap
        in the desynchronized case.

		\begin{figure*}[t!]
			\centering
			\input{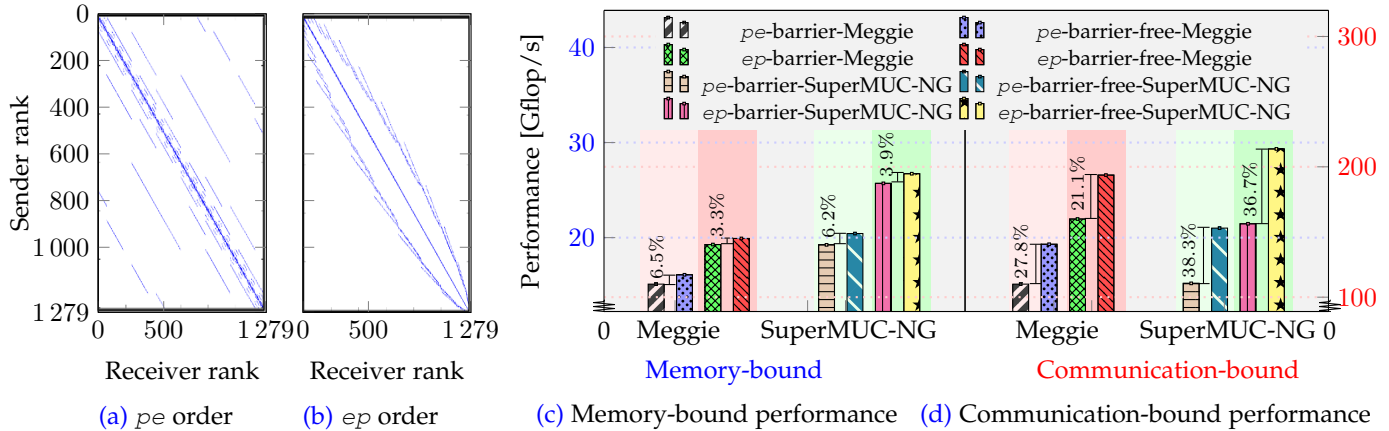}%
			\caption{{\textcolor{blue}{(a-b)}} Communication
				topology of \texttt{HHQ-small-order\emph{pe}} and
				\texttt{HHQ-small-order\emph{ep}} Hamiltonian
				matrices.  {\textcolor{blue}{(c)}} Performance (blue
				$y$-axis) and speedup for barrier-free
				(desynchronized) execution in the memory-bound case
				($60$ processes on Meggie and $96$ processes on
				SuperMUC-NG) and in non-split mode. The
				percentage increments denote the speedup of the
				no-barrier versions.  {\textcolor{blue}{(d)}} Same
				data but for the communication-bound case with
				performance on the red $y$-axis ($1280$ processes on
				Meggie and $1296$ processes on SuperMUC-NG).
				%
			}
			\label{fig:SPMV2} 
		\end{figure*}
        Note that the ``slimmer'' \emph{ep} matrix with its smaller
        communication radius supports stronger desynchronization due
        to a lower idle wave velocity~\cite{AfzalHW20,AfzalHW2021}.
        This effect is counteracted, however, by the smaller absolute
        communication overhead of \emph{ep}, which is why no clear
        advantage of \emph{ep} in terms of overlap can be observed in
        Fig.~\ref{fig:SPMV}(e)\@. Note also that particular
          process counts can interact with the inherent structure
          of the matrix, which leads to more or less favorable
          communication topologies and adds extra variation to the
          scaling behavior.
        The general trend is
        similar but less pronounced on the Meggie system, as shown by
        the black markers in Fig.~\ref{fig:SPMV}. This can be
        attributed to the larger fraction of cores needed per ccNUMA
        domain to achieve memory bandwidth saturation compared to
        SuperMUC-NG.
        
        In this experiment, the matrices were large
        enough to keep the execution memory bound even at large
        scale, which made memory bandwidth the relevant bottleneck
        for desynchronization. The \texttt{HHQ-small} matrices in Table~\ref{tab:mat}
        fit into the aggregate last-level cache (LLC) on 23 nodes and
        beyond (for Meggie)
        and 18 nodes (for SuperMUC-NG), respectively. The LLC shows
        much better bandwidth scalability than the memory interface,
        so the bottleneck shifts to the network communication
        for larger node counts. In Fig.~\ref{fig:SPMV2} we show
        performance and speedup results for the small matrices
        on small (memory bound, (c))
        and large (network bound (d)) numbers of nodes on the two test
        systems. As before, the actual barrier duration has been
        subtracted from the execution time of the version with
        barriers so that the speedup observed when removing the
        barrier can be attributed to desynchronization.

        We concentrate
        on the non-split variant here.
        For small numbers of processes (Fig.~\ref{fig:SPMV2}(c)),
        the speedups are small,
        which is expected because the memory bandwidth is the bottleneck
        and the communication overhead is small so that there
        is little opportunity for overlapping.
          For the in-memory case, the small \emph{ep} matrix shows the
          same desynchronization speedup of $3.9$\,\% as its large
          counterpart.  The more communication-intensive \emph{pe}
          matrix, on the other hand, shows an extra speedup of
          $1.45$\,\% (this data is not contained in the figure).  The
          behavior persists in similar ways on both systems.  In the
        network-bound case (Fig.~\ref{fig:SPMV2}(d)), however, the
        now-dominant communication can overlap with code execution,
        which yields speedups of 38\% and 37\% in the \emph{pe} and
        \emph{ep} cases, respectively. We attribute the minor
        difference in behavior between the two matrices to the small
        message sizes, which make most of the point-to-point
        communication latency bound.  
          Compared to the large-matrix cases, the in-cache
          execution leads to lower speedup by desynchronization
          ($4.4$\,\% for \emph{ep} and $8$\,\% for \emph{pe}).
          On the Meggie system, the lower CER
          causes an additional boost by $15.6$\,\% (\emph{ep})
          and $10.5$\,\% (\emph{pe}), respectively.  

        In summary, our SpMVM experiments have shown that significant
        performance speedups can be obtained via desynchronization
        when the execution is limited by memory bandwidth or
        communication and synchronizing collectives (i.e., barriers or
        collectives with synchronizing implementations) between
        back-to-back SpMVMs are removed. Note that we relied on the
        natural irregularities of the sparse matrices to destabilize
        the lock-step pattern; no explicit noise injection was
        required.  \highlight{\emph{Key takeaway}: In MPI-parallel
          SpMVM, speedup by automatic
          overlap of communication and computation is facilitated
          by (i) a larger communication overhead, which is connected with
          a more spread nonzero distribution in the matrix, (ii) a slow
          idle wave propagation speed, which is caused by a low matrix
          bandwidth, and (iii) a slow synchronized baseline that uses
          the simple non-split communication scheme.}

	\begin{algorithm*}[t]
		\centering
		\caption{{Structure of the MPI-parallel \textsc{chebFD($ H\texttt{,} \vec U\texttt{,} \vec W\texttt{,}\vec X$)} implementation. \textsc{Non-split}, \textsc{split-wait} and \textsc{pipeline} modes are implementation alternatives.}}
		\centering
\hspace{-1.7em}
\begin{minipage}{0.18\textwidth}%
	\begin{frame}
		\centering
		\tikzset{every shadow/.style={fill=none,scale=0}}
		\tikzset{description/.append style={top color=\col,bottom color=\col}}
		\tikzset{priority arrow/.append style={rotate=180,anchor=0,xshift=5}}
		\smartdiagramset{
			set color list={orange!10,pink!30,blue!10,green!5},
			uniform connection color=true,
			back arrow disabled=true,
			module shape=rectangle, 
			border color=none,
			module minimum height=0.5cm,
			description text width=3.2cm,
			description title width=1cm,
			descriptive items y sep=39pt,
			font=\Huge\sffamily\bfseries, 
			priority arrow width=0.3cm,
			priority arrow height advance=1.05cm,
			priority arrow head extend=0.04cm,
		}
		\smartdiagram[priority descriptive diagram]{ 
			Main loop: \textsc{Non-split}/\textsc{Split}/\textsc{Pipeline}~mode,
			\textsc{baxpy()} and \textsc{bscal()} \\ $\mathrlap{\vec X}\phantom{\vec W} \gets g_0 c_0 \vec X + g_1 c_1
			\vec U + g_2 c_2 \vec W$,
			twice \textsc{spMMV()}\\$\mathrlap{\vec U}\phantom{\vec W} \gets (\alpha\vec H + \beta) \vec X
			\texttt{,}$\\
			$\vec W \gets 2 (\alpha\vec H + \beta) \vec U - \vec X$,
			define vector blocks\\  $\vec U\texttt{,} \vec W\texttt{,} \vec X$
		}
	\end{frame}
\end{minipage}
\hspace{8em}
\vspace{-1.3em}
\begin{minipage}{0.21\textwidth}%
	\vspace{-1.2em}
	\textbf{\textsc{Non-split} mode}
	
	\textbf{for} k $= 0 : \frac{n_{s}}{n_{b}}-1$  \textbf{do} \\
	\textbf{for} p $= 0 : n_{p}-1$ \textbf{do}\\
	$\mbox{swap} ( \vec W ,  \vec U )$ \;
	\tikzmk{A}
	$\mbox{comm\_init\_finalize} (\vec U)$ \;	
	\tikzmk{B}
	\boxit{red!40}	
	\tikzmk{A}
	$\vec W \gets 2 (\alpha\vec H + \beta) \vec U -  \vec W$\;
	\tikzmk{B}
	\boxit{yellow!40}
	\tikzmk{A}
	$\mathrlap{\vec \eta_p}\phantom{\vec X} \gets \langle \vec W, \vec U \rangle$ \;
	$\mathrlap{\vec \mu_p}\phantom{\vec X} \gets \langle \vec U, \vec U \rangle$ \;
	\tikzmk{B}
	\boxit{green!20}		
	\tikzmk{A}
	$\mathrlap{\vec X}\phantom{\vec W} \gets \vec X + g_p c_p \vec W$ \;
	\tikzmk{B}
	\boxit{blue!15}
	\textbf{end for}\\
	\textbf{end for}
\end{minipage}%
	\smash{\hspace{0em}\raisebox{-3.2em}{\rotatebox{90}{\text{kernel}}}\hspace{-1.6em}\raisebox{-2.2em}{$\left.\begin{array}{@{}c@{}}\\{}\\{}\\{}\\{}\end{array}\right\}$}}
\hspace{1em}
\begin{minipage}{0.19\textwidth}%
	\vspace{-1.5em}
	\textbf{\textsc{Split-wait} mode}
	
	\textbf{for} k $= 0 : \frac{n_{s}}{n_{b}}-1$  \textbf{do} \\
	\textbf{for} p $= 0 : n_{p}-1$ \textbf{do}\\
	$\mbox{swap} ( \vec W ,  \vec U )$ \;
	\tikzmk{A}
	$\mbox{comm\_int} (\vec U)$ \;
	\tikzmk{B}
	\boxit{red!40}	
	\tikzmk{A}
	$\mbox{local\_kernel}$ \;
	\tikzmk{B}
	\boxit{cyan!60}
	\tikzmk{A}
	$\mbox{MPI\_wait()}$ \;
	\tikzmk{B}
	\boxit{red!40}
	\tikzmk{A}
	$\mbox{remote\_kernel}$ \;
	\tikzmk{B}
	\boxit{cyan!60}
	\textbf{end for}\\
	\textbf{end for}
\end{minipage}%
\hspace{0.5em}
\begin{minipage}{0.19\textwidth}%
	\vspace{-1.5em}
	\textbf{\textsc{pipeline} mode}
	
	\textbf{for} p $= 0 : n_{p}-1$ \textbf{do}\\
	$\mbox{swap} ( \vec W ,  \vec U )$ \;
	\tikzmk{A}
	$\mbox{comm\_init} (\vec U_0)$ \;
	$\mbox{MPI\_wait()}$ \;
	\tikzmk{B}
	\boxit{red!40}
	\textbf{for} k $= 0 : \frac{n_{s}}{n_{b}}-2$ \textbf{do} \\
	\tikzmk{A}
	$\mbox{comm\_init} (\vec U_1)$ \;
	\tikzmk{B}
	\boxit{red!40}		
	\tikzmk{A}
	$\mbox{kernel}$$_k$$\mbox{()}$ \;
	\tikzmk{B}
	\boxit{cyan!60}
	\tikzmk{A}
	$\mbox{MPI\_wait()}$ \;
	\tikzmk{B}
	\boxit{red!40}
	\textbf{end for}\\
	\tikzmk{A}
	$\mbox{kernel}$$_{\frac{n_{s}}{n_{b}}-1}$$\mbox{()}$ \;
	\tikzmk{B}
	\boxit{cyan!60}
	\textbf{end for}	
\end{minipage}%
		\label{alg:chebfd}
	\end{algorithm*}
	\begin{table}[t]
		\centering
		\caption{Key specifications of Hamiltonian matrices for the ChebFD application.} 
		\label{tab:matChebFD}
		\begin{adjustbox}{width=0.49\textwidth}
			\begin{threeparttable}
				\setlength\extrarowheight{-0.7pt}
\setlength\tabcolsep{2pt}
\arrayrulecolor{blue}
\begin{tabular}[fragile]{c>{~}wxyyyz}
	\toprule
	\rowcolor[gray]{0.9}
	{Matrix}&{Traits}\mbox{$^\mathsection$}& {Data-type}&{$n_\mathrm{r}=n_\mathrm{c}$}	&{$n_\mathrm{nz}$}	&{$n_\mathrm{nzpr}$}	&{Size [\si{\giga\bytes}]\mbox{$^\ddag$}}\\
	\midrule
	\textsc{Topi-enh}&$m_x$-$m_y$-$m_z$-$128$-$128$-$64$&complex double&268435456&3487563776&13&70.8\mbox{$^\ddag$}\\
	\textsc{Spin$26$}\mbox{$^\star$}&
	$26$-$13$-$1$&double&10400600&145608400&14&1.8\\
	\textsc{Spin$28$}\mbox{$^\star$}&
	$28$-$14$-$1$&double&40116600&601749000&15&7.4\\
	\textsc{Spin$30$}\mbox{$^\star$}&
	$30$-$15$-$1$&double&155117520&2481880320&16&30.4\\
	\bottomrule
\end{tabular}

				\begin{tablenotes}[flushleft]
					\normalsize
					\item \mbox{$^\mathsection$} Traits for the \textsc{Topi-enh} matrix represent the $m_x$-$m_y$-$m_z$-$n_x$-$n_y$-$n_z$, while for the \textsc{spin} matrices mark
					$n_{up}$-disorder-seed.
					\item \mbox{$^\ddag$}
					Eight (sixteen) byte for double (complex double) precision numbers of matrix entries, and four-byte indexing for 32-bit integers are considered.
					\item \mbox{$^\star$}
					\textsc{Spin} matrices comprising periodic boundary conditions with \emph{upper} count are equal to the half of the \emph{lower}.
				\end{tablenotes}
			\end{threeparttable}
		\end{adjustbox}
	\end{table}
	\begin{table*}[t!]
		\centering
		\caption{{\textcolor{blue}{(a)}} ChebFD message profile of topological insulator
			(\textsc{Topi-ehn}) matrices of same problem sizes. It encompasses nine
			domain configurations (\textsc{M$1$}--\textsc{M$9$}) with $n_\mathrm{nodes}$
			that facilitates a good mixture of eager and rendezvous messages.
			{\textcolor{blue}{(b)--(c)}} Communication matrices of {Spin} matrices of multiple sizes.}
		\label{fig:ChebFD_TopiDomain}
		\hspace{-2em}
\vspace{-1em}
\begin{minipage}[c]{0.64\textwidth}
\vspace{-1em}
\resizebox{\textwidth}{!}{%
	\begin{tabular}[fragile]{cwxz}
		\toprule
		\rowcolor[gray]{0.9}&{$m_x$-$m_y$-$m_z$}&{Communication distances ($n_\mathrm{nodes}=64$)}&{Message sizes [\si{\kilo\bytes}] ($n_\mathrm{nodes}=64$)}\\
		\midrule
		{\textsc{M$1$}}&{$1$-$1$-$n_\mathrm{nodes}$}&{$\pm1,-20,-19$}&{$1050,105,0.128$}\\
		{\textsc{M$2$}}&{$1$-$n_\mathrm{nodes}$-$1$}&{$\pm1,\pm20,\pm19$}&{$1050,48,8$}\\
		{\textsc{M$3$}}&{$n_\mathrm{nodes}$-$1$-$1$}&{$\pm1$}&{$1050$}\\
		{\textsc{M$4$}}&{$1$-$\sqrt{n_\mathrm{nodes}}$-$\sqrt{n_\mathrm{nodes}}$}&{$\pm1,-20,\pm160,\pm159,-19$}&{$1050,105,48,8,0.128$}\\
		{\textsc{M$5$}}&{$\sqrt{n_\mathrm{nodes}}$-$1$-$\sqrt{n_\mathrm{nodes}}$}&{$\pm1,-20,-19$}&{$1050,105,0.128$}\\
		{\textsc{M$6$}}&{$\sqrt{n_\mathrm{nodes}}$-$\sqrt{n_\mathrm{nodes}}$-$1$}&{$\pm1,-20,-140,-19,-141$}&{$1050,48,48,8,8$}\\
		{\textsc{M$7$}}&{$2$-$4$-$\sqrt{n_\mathrm{nodes}}$}&{$\pm1,-20,-480,-160,-481,-159,-19$}&{$1050,105,48,48,8,8,0.128$}\\
		{\textsc{M$8$}}&{$2$-$\sqrt{n_\mathrm{nodes}}$-$4$}&{$\pm1,-20,-560,-80,-561,-79,-19$}&{$1050,105,48,48,8,8,0.128$}\\
		{\textsc{M$9$}}&{$\sqrt{n_\mathrm{nodes}}$-$2$-$4$}&{$\pm1,-20,-80,-79,-81,-19$}&{$1050,105,96,8,8,0.128$}\\
		\bottomrule
\end{tabular}}
	\begin{tikzpicture}		
		\node [] at (10,-1.4){\textcolor{blue}{(a)}~\textsc{Topi-ehn}};
	\end{tikzpicture}
\end{minipage}
\begin{minipage}[c]{0.3\textwidth}
\vspace{-1em}
	\begin{subfigure}[t]{0.15\textwidth}
		\begin{tikzpicture}
			\put(-0.4,0) {\includegraphics[width=0.09\textheight,height=0.1\textheight]{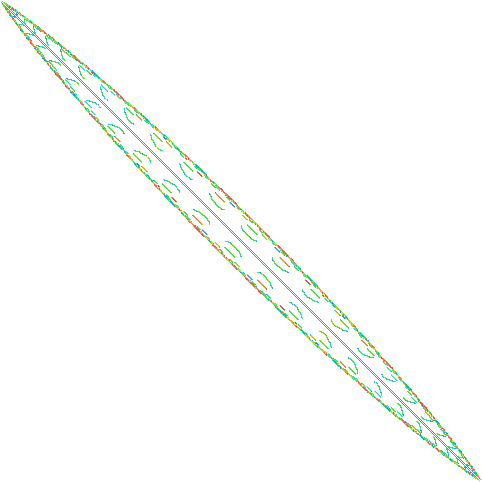}} 
			\begin{axis}[trim axis left, trim axis right, scale only axis,
				width = 0.09\textheight,
				height = 0.1\textheight,
				ylabel = {Sender rank},
				xlabel = {Receiver rank},
				y label style={at={(0.08,0.5)}},
				xmin=0, xmax=1284,
				ymin=0, ymax=1279,
				xtick={0,500,1279},
				y dir=reverse,
				ytick={0,200,400,600,800,1000,1279},
				axis on top,
				]
			\end{axis}
			\node [] at (0.9,-1.4){\textcolor{blue}{(b)}~\textsc{Spin26/30}};
		\end{tikzpicture}
	\end{subfigure}
	\hspace{1em}
	\begin{subfigure}[t]{0.15\textwidth} 
		\begin{tikzpicture}
			\put(-0.4,0) {\includegraphics[width=0.09\textheight,height=0.1\textheight]{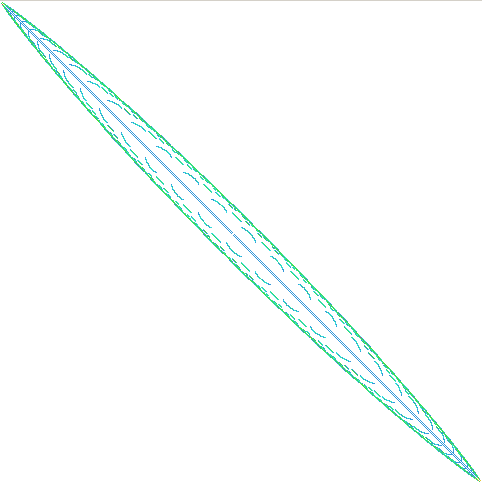}} 
			\begin{axis}[trim axis left, trim axis right, scale only axis,
				width = 0.09\textheight,
				height = 0.1\textheight,
				xlabel = {Receiver rank},
				y label style={at={(-0.5,0.5)}},
				xmin=0, xmax=1284,
				ymin=0, ymax=1279,
				xtick={0,500,1279},
				y dir=reverse,
				ytick={0,200,400,600,800,1000,1279},
				yticklabels=\empty,
				axis on top,
				]
			\end{axis}
			\node [] at (0.9,-1.4){\textcolor{blue}{(c)}~\textsc{Spin28}};
		\end{tikzpicture}
	\end{subfigure}
\end{minipage}
	\end{table*}
\subsection{Chebyshev {F}ilter {D}iagonalization ({ChebFD})}\label{sec:chebFD}		
The \ac{ChebFD} is a polynomial filtering algorithm that is popular
in quantum physics and chemistry. It allows to compute parts of eigenvalue spectra
of large sparse matrices and is amenable to multiple node-level and
communication optimizations \cite{pieper2016high,kreutzer2018chebyshev}.
For example, a blocking parameter (number of block vectors $n_{b}$)
enables flexible tuning of the code balance of the main iteration loop; larger
block size causes lower code balance.
			
\subsubsection{Implementation}
Our open-source\footnote{Available as part of the GHOST package at \url{https://bitbucket.org/essex/ghost}}
implementation of \ac{ChebFD} is built with the GHOST~\cite{GHOST2017} building block library using
tailored kernels and \SI{64}{\bits} global and \SI{32}{\bits} local index size, respectively.
It employs the standard CRS sparse matrix data format and uses row-major ordering within a block of $n_b$ vectors to facilitate SIMD vectorization. 
The algorithm contains a sparse matrix-multiple-vector multiplication (SpMMV) and a series of BLAS-1 vector
operations. Non-blocking communication is performed via
\texttt{MPI\_Isend}/\texttt{MPI\_Irecv}/\texttt{MPI\_Waitall} sequences. Asynchronous progress was disabled in the Intel MPI library as well as in GHOST. The code supports
  hybrid MPI+OpenMP parallelization; unless otherwise noted, we use the pure MPI version here.

We compare three communication schemes:
\begin{enumerate}[\setlength\topsep{0pt}]
\item \textbf{\textsc{Non-split} mode}: blocking MPI communication, followed by computation.
  It performs computation of ChebFD polynomials to a block $\vec U$ of
  $n_{b}$ vectors at a time.
\item \textbf{\textsc{Split-wait} mode}:
  naive implicit overlapping of non-blocking MPI communication of the non-local vector elements
  with local computations. Only after completing the outstanding receives via \verb.MPI_Wait.
  can the remote part of the kernel be done. It increases the main memory data traffic
  since the local result vector must be updated twice.
\item \textbf{\textsc{pipeline} mode}: 
  pipelined asynchronous non-blocking MPI communication with the subspace blocking scheme,
  which does not require any extra memory traffic.  If $n_s$ is the subspace size,
  for sufficiently large $\frac{n_{b}}{n_{s}}$ this scheme enables explicit effective overlap of 
  computation on the current subblock (local\_kernel) with the communication needed for the next
  sub-block. Details can be found in~\cite{kreutzer2018chebyshev}.
\end{enumerate}
The polynomial filter degree (we use $n_{p}=500$ here, which is a relevant
value for practical applications) applies independently to
all search vectors $n_{s}$.  In the algorithm, $n_\mathrm{iter}$ is the
number of iterations; the number of sought inner eigenvalues of a topological insulator
($n_{s} = 128$) is taken to be a multiple of block vector size ($n_{b}
= 2$ or $n_{b} = 32$) for simplicity.  In the implementation,
a single, fused, MPI-parallel \textsc{ChebFD()} kernel (marked in the algorithm)
facilitates cache reuse. Moreover, the global reduction needed for computation of the
polynomial filter coefficients $\vec \eta_p$ and $\vec \mu_p$ can
be postponed until after the iteration loop, which eliminates all (possibly) synchronizing
collectives; the communication topology is thus entirely determined by the matrix structure.
The sparse matrices and the code balance of the algorithm
(optimistically\footnote{``Optimistically'' means here that one assumes the minimum possible data transfer, i.e., each data element is only loaded once and then reused from cache as often as necessary.} $(260/n_b+80)/146\,\si{\byte \per flop}$ for double complex data and $ (48/n_b+40)/19\,\si{\byte \per flop} $ for double precision real data)
were thoroughly investigated in~\cite{pieper2016high}.
\begin{figure}[t]
	\centering
\hspace{-0.7em}
\begin{minipage}[t]{0.23\textwidth}
	\begin{tikzpicture}
		\pgfplotstableread{figures/fig_ChebFD/Topi-1-1-1-128-128-64_ns128_nb2_Vector_chunkheight1_implAVX_aligned.txt}\Ablocksz;
		\pgfplotstableread{figures/fig_ChebFD/Topi-1-1-1-128-128-64_ns128_nb32_Vector_chunkheight1_implAVX_aligned.txt}\Bblocksz;
		\begin{axis}[trim axis left, trim axis right, scale only axis,
			axis background/.style={fill=white!95!black},
			width=0.8\textwidth,height=0.13\textheight,
			xlabel = {Threads per socket},
			ybar, 
			ybar legend, 
			bar width=0.8mm,
			xmin=0,
			ymin=0,
			ymax=70,
			xmax=11,
			xtick pos=left,
			ytick pos=left,
			y label style={at={(0.12,0.5)},font=\footnotesize},
			x label style={font=\footnotesize},
			x tick label style={font=\footnotesize},
			ylabel = {Performance [\si{\giga \flop / \second}]}, 
			xtick={0,2,4,6,8,10},
			xticklabels={0,2,4,6,8,10},
			ytick={0,10,20,30,40,50,60},
			ymajorgrids,
			legend columns = 1, 
			legend style = {
				nodes={inner sep=0.04em},
				draw=none,
				font=\scriptsize,
				cells={align=left},
				anchor=east,
				at={(0.65,0.89)},
				fill=white!95!black,
				/tikz/column 1/.style={column sep=5pt,},
			},
			]		
			
			\addplot[ postaction={
				pattern=horizontal lines
			}, fill={myred!50!white}]
			table
			[
			x expr=\thisrow{Cores}, 
			y expr=\thisrow{GHOST_perf[GF/s]},
			]{\Bblocksz};
			\addlegendentry{~$n_\mathrm{b}=32$}	  			
		\end{axis}

		\begin{axis}[trim axis left, trim axis right, scale only axis,
			xshift=2.75pt,
			width=0.8\textwidth,height=0.13\textheight,
			ybar, 
			ybar legend, 
			bar width=0.8mm,
			xmin=0,
			ymin=0,
			ymax=70,
			xmax=11,
			ytick={},
			ymajorgrids,
			axis lines=none, 
			legend columns = 1, 
			legend style = {
				nodes={inner sep=0.04em},
				draw=none,
				font=\scriptsize,
				cells={align=left},
				anchor=east,
				at={(0.57,0.75)},
				fill=white!95!black,
				/tikz/column 1/.style={column sep=5pt,},
			},
			]		
			\addplot[ postaction={
				pattern=north east lines
			}, fill={myblue!50!white}]
			table
			[
			x expr=\thisrow{Cores}, 
			y expr=\thisrow{GHOST_perf[GF/s]},
			]{\Ablocksz};
			\addlegendentry{~$n_\mathrm{b}=2$}	
		\end{axis}
		\node [] at (1.5,-1.4){\textcolor{blue}{(a)}~\textsc{Topi-ehn} matrix};
	\end{tikzpicture}
\end{minipage}
\hspace{-2em}
\begin{minipage}[t]{0.23\textwidth}
	\begin{tikzpicture}
		\pgfplotstableread{figures/fig_ChebFD/Spin-26-13-1_ns128_nb2_Vector_chunkheight1_implAVX_aligned.txt}\Ablocksz;
		\pgfplotstableread{figures/fig_ChebFD/Spin-26-13-1_ns128_nb32_Vector_chunkheight1_implAVX_aligned.txt}\Bblocksz;
		\begin{axis}[trim axis left, trim axis right, scale only axis,
						axis background/.style={fill=white!95!black},
			width=0.8\textwidth,height=0.13\textheight,
			xlabel = {Threads per socket},
			ybar, 
			ybar legend, 
			bar width=0.8mm,
			xmin=0,
			ymin=0,
			ymax=30,
			xmax=11,
			xtick pos=left,
			ytick pos=left,
			x label style={font=\footnotesize},
			x tick label style={font=\footnotesize},
			xtick={0,2,4,6,8,10},
			xticklabels={0,2,4,6,8,10},
			ytick={0,10,20,30,40,50,60},
			ymajorgrids,
			legend columns = 1, 
			legend style = {
				nodes={inner sep=0.04em},
				draw=none,
				font=\scriptsize,
				cells={align=left},
				anchor=east,
				at={(0.67,0.89)},
				fill=white!95!black,
				/tikz/column 1/.style={column sep=5pt,},
			},
			]		
			
			\addplot[ postaction={
				pattern=horizontal lines
			}, fill={myred!50!white}]
			table
			[
			x expr=\thisrow{Cores}, 
			y expr=\thisrow{GHOST_perf[GF/s]},
			]{\Bblocksz};
			\addlegendentry{~$n_\mathrm{b}=32$}	  			
		\end{axis}
		
		\begin{axis}[trim axis left, trim axis right, scale only axis,
			xshift=2.8pt,
			width=0.8\textwidth,height=0.13\textheight,
			ybar, 
			ybar legend, 
			bar width=0.8mm,
			xmin=0,
			ymin=0,
			ymax=30,
			xmax=11,
			ytick={},
			ymajorgrids,
			axis lines=none, 
			legend columns = 1, 
			legend style = {
				nodes={inner sep=0.04em},
				draw=none,
				font=\scriptsize,
				cells={align=left},
				anchor=east,
				at={(0.59,0.75)},
				fill=white!95!black,
				/tikz/column 1/.style={column sep=5pt,},
			},
			]		
			\addplot[ postaction={
				pattern=north east lines
			}, fill={myblue!50!white}]
			table
			[
			x expr=\thisrow{Cores}, 
			y expr=\thisrow{GHOST_perf[GF/s]},
			]{\Ablocksz};
			\addlegendentry{~$n_\mathrm{b}=2$}	
		\end{axis}
		\node [] at (1.5,-1.4){\textcolor{blue}{(b)}~\textsc{Spin26} matrix};
	\end{tikzpicture}
\end{minipage}
	\caption{Single-socket performance scaling of ChebFD with OpenMP on a contention domain of the Meggie system for the block vector sizes of $2$ and $32$ and the
		{\textcolor{blue}{(a)}} topological insulator matrix (1.76 speed up at single thread) and
		{\textcolor{blue}{(b)}} \textsc{Spin26} matrix (2.89 speed up at single thread owing to efficient data accesses), respectively.}
	\label{fig:chebFD_socket}
\end{figure}
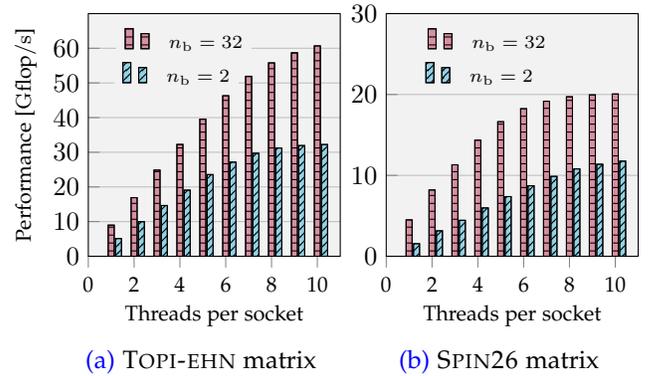
\subsubsection{Test matrices}

Two types of test matrices and scaling scenarios were considered for ChebFD: 
weak scaling for a topological insulator problem (\textsc{Topi-enh}) ~\cite{sitte2012topological} and
strong scaling for a spin system (\emph{Spin}). For \textsc{Topi-enh} (see Table~\ref{tab:matChebFD}), the Hamiltonian
matrix emerges from a three-dimensional mesh with four \ac{DOFS} per
mesh point and thus exhibits a rather regular, stencil-like structure. In order
to study different communication topologies, we chose a local
(per-node) problem size of $n_x\times n_y\times n_z=128^2\times 64$ and a global size
of $n_xm_xn_ym_yn_zm_z$, where the $m_i$ are the number of nodes in each
Cartesian dimension. Each matrix row has 13 complex double-precision nonzero entries,
leading to a matrix size of $(16+4)\,\bytes \times 128 \times 128 \times 64 \times 13=273\,\MB$, which much larger than the available LLC on the benchmark platforms.
Periodic boundary conditions in the $x$ and
$y$ directions lead to outlying diagonals in the matrix corners.

The real-valued \textsc{Spin} matrix is used in three different, fixed sizes (see Table~\ref{tab:matChebFD}) and has a structure that leads to more communication
  overhead and a larger impact of memory latency on the node-level performance.
  The socket-level performance scaling data (using OpenMP)
  in Fig.~\ref{fig:chebFD_socket} reveals
  interesting differences between the two matrices: Although increasing the
  block size from $n_b=2$ to $n_b=32$ improves the performance in both cases
  as expected from the reduced code balance, the impact on the scaling behavior is
  different: While the \textsc{Spin-26} matrix starts off with very low performance
  on a single thread with $n_b=2$ and thus shows good scaling in this case
  (optimistic upper bandwidth limit at 14.8\,\GFS), the \textsc{Topi-enh} matrix shows strong
  bandwidth saturation and achieves 90\% of the optimistic maximum of 34.8\,\GFS\
  already at eight cores. At $n_b=32$, the code balance is strongly reduced
  in both cases, so one expects weaker saturation as the pressure on the
  memory interface is lowered and in-cache effects become more
  prominent~\cite{pieper2016high,Kreutzer:2015}. While this can be clearly observed
  in case of \textsc{Topi-enh}, saturation actually becomes stronger for
  \textsc{Spin-26}. This is rooted in a better utilization of the memory interface
  and a lower impact of latency due to the vector blocking technique.

  All matrices are generated on the fly using the
  \texttt{ESSEX\_PHYSICS} library\footnote{The \texttt{ESSEX\_PHYSICS}
    library is a open-source software of ESSEX project, available for
    download at
    \url{https://bitbucket.org/essex/physics/src/master/}.}.
  Table~\ref{tab:matChebFD} illustrates the key specifications of both
  types of matrices.  

\begin{figure*}[t]
	\centering
	\input{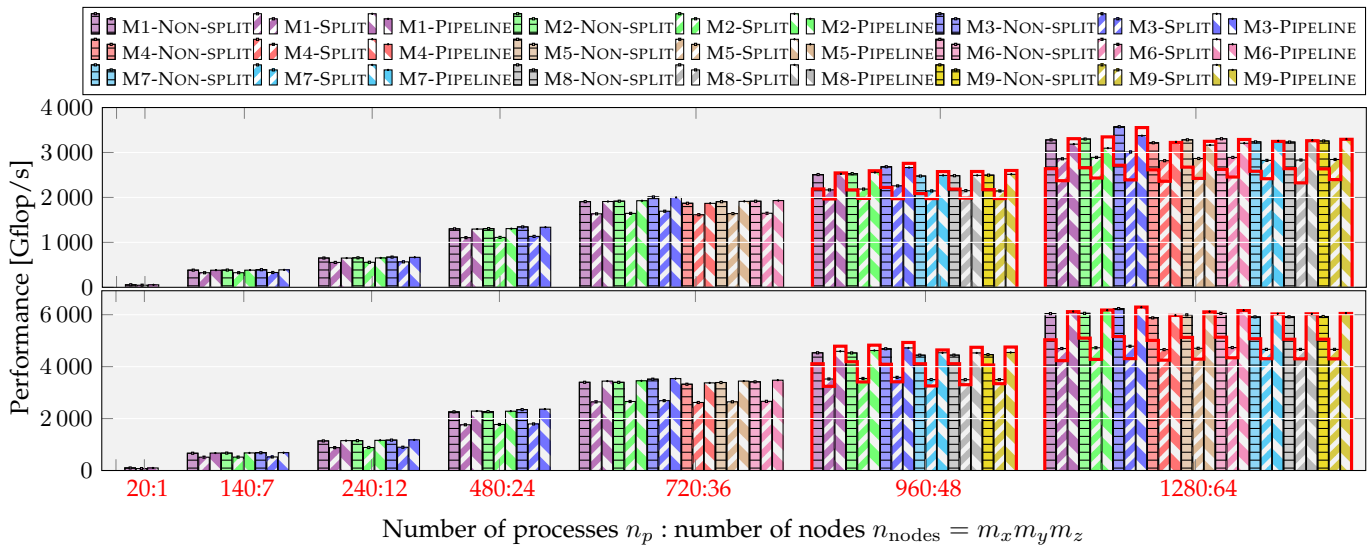}%
	\caption{Weak scaling performance of ChebFD with \textsc{Topi-enh} matrices
		in nine domain decompositions
		(\textsc{M$1$}--\textsc{M$9$} in different colors) with three communication schemes
		(\textsc{Non-split} in horizontal lines, \textsc{Split-wait} in north east hatch lines, \textsc{Pipeline} in north west hatch lines) 
		on the Meggie cluster.  The number of MPI processes and
		compute nodes are shown on the $x$-axis.
		Top: block vector size of $n_b=2$, bottom: $n_b=32$.
		The red line indicates performance with a barrier in each $p$ iteration (see text). }
	\label{fig:ChebFD_Topi} 
\end{figure*}

\subsubsection{Decomposition, communication schemes and block vector sizes}

\paragraph*{\textbf{\textsc{Topi-enh} matrices}}
Figure ~\ref{fig:ChebFD_Topi} shows the weak scaling performance of
ChebFD with the \textsc{Topi-enh} case in nine domain decomposition
variants and three communication modes. While the bars show the observed performance with
barrier-free code, the red line denotes the performance with an
explicit barrier added at each new polynomial degree $p$ (specifically
when the $\vec W$ and $\vec V$ vectors are swapped). As expected, the
pipelined mode exhibits no noticeable performance hit from the barrier
because the vector swap occurs between the $p$ and $k$ loops.
The communication topology depends on the domain decomposition, as
do the communication data paths used (intra- vs.\ internode). Hence, we expect
a significant dependence of the performance on the decomposition.
Unlike in \ac{SpMVM}, where the matrix data dominates the code balance,
ChebFD has a much stronger dependence on the vector data;
consequently, the split-wait variant shows the worst performance among all
schemes and decompositions due to extra memory traffic.  Favorable configurations
for the sync-free code are (in order of descending performance):
$\{\textsc{M$3$},\textsc{M$2$},\textsc{M$6$},\textsc{M$5$},\textsc{M$1$},\textsc{M$9$},\textsc{M$7$},\textsc{M$8$},\textsc{M$4$}\}$,
which is also roughly in rising idle wave speed order. With explicit barriers,
on the other hand, \textsc{M$7$} is the worst configuration and
\textsc{M$8$} is among the top performers.
Configuration \textsc{M$3$} uses a one-dimensional decomposition
and thus has an unfavorable communication pattern.
However, it features only direct next-neighbor communication,
which leads to the slowest possible idle wave speed and thus
the highest potential for desynchronization.

\textsc{M$4$} the worst case among all studied decomposition
variants, and its CER is even larger than that of \textsc{M$3$}.
For instance, at the blocking vector size of $32$, we observe
with the synchronized non-split variant
\{$T_{exec}$ [ms], $T_{comm}$ [ms], CER\} = \{214.6,30,0.14\} for
\textsc{M$3$} and \{258.6,43,0.17\} for \textsc{M$4$}.
Thus, the configuration \textsc{M$4$} spends more time
communicating, while \textsc{M$3$} has the
slowest idle wave; both affect the overall performance.  In
\{\textsc{Non-split}, \textsc{pipeline}, \textsc{Split-wait}\} mode,
the performance increment between the worst \textsc{M$4$} and best
\textsc{M$3$} corner cases is \{11.1, 9, 6.9\}\,\% with $n_b=2$ and
\{6.1, 5.1, 2.8\}\,\% with $n_b=32$ at $1280$ MPI processes on Meggie.
The $n_b=32$ cases
exhibit a stronger slowdown from the non-split to the split version due to the
dominant data traffic from the vector blocks. The pipeline version is always
better than non-split in the synchronized scenario, while in the
naturally desynchronized case, the non-split version is generally on
par or even better for the more saturating case.
Contrarily, the non-split version suffers a higher performance hit
at $n_b=2$ with synchronizing barriers in place.\smallskip
\highlight{\emph{Key takeaway}:
  The domain decomposition that allows 
		lower idle wave speeds
		is better suited for automatic communication overlap.}\smallskip

\begin{figure*}[t]
  \centering
  \input{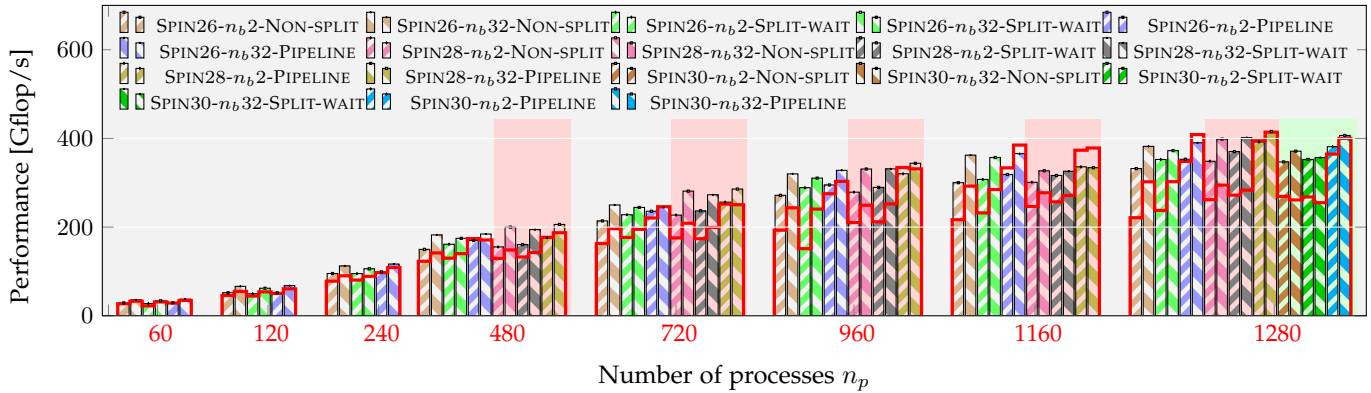}%
  \caption{Strong scaling performance of ChebFD with
    \textsc{Spin} matrices of diverse sizes (\textsc{Spin26} in gray, \textsc{Spin28} in red, \textsc{Spin30} in green background) using various communication schemes (\textsc{Non-split}, \textsc{Split-wait}, \textsc{Pipeline} in different colors) with block vector sizes of $n_b=2$ (north east hatch lines) and $n_b=32$ (north west hatch lines) on the Meggie system.
    The red line indicates performance with a barrier in each $p$ iteration. 
  }
  \label{fig:ChebFD_Spin} 
\end{figure*}
\paragraph*{\textbf{\textsc{Spin} matrices}} Figure
\ref{fig:ChebFD_Spin} shows strong scaling results for the
\textsc{Spin} matrices. Due to the working set size, \textsc{Spin-28}
can only be used on 36 nodes and more, and \textsc{Spin-30} requires
64 nodes at least.  Aggregate LLC sizes of
$\{0.05,0.35,0.6,1.2,1.8,2.4,3.2\}$\,\GB\ are available for
$\{60,120,240,480,720,960,1280\}$ MPI processes on the Meggie system.
As a consequence, contrary to the \textsc{Spin28} and
\textsc{Spin30} matrices, ChebFD with the \textsc{Spin26} matrix
starts to be cache bound from 720 processes up.
	
In \textsc{Non-split} mode and on 64 nodes on Meggie, we observe
\{$T_{exec}$ [ms], $T_{comm}$ [ms], CER\} = 
\{0.6,0.41,0.68\}, \{2.38,1.86,0.78\} and \{10.2,7,0.69\} for
\textsc{Spin26}, \textsc{Spin28} and \textsc{Spin30}, respectively.
Similarly, the range of \ac{P2P} message sizes is
\{$V_{min}$ [\si{\bytes}], $V_{max}$ [\si{\kilo\bytes}] (red)\} = \{16,130\}
(\textsc{Spin26}), \{64,501.5\} (\textsc{Spin28}) and \{32,1939\}
(\textsc{Spin30}).  These matrices cause significantly more communication
overhead than \textsc{Topi-enh}, which leads to more opportunity
for desynchronized execution and communication overlap in the \textsc{Non-split}
case.
The case with $n_b=32$ shows stronger socket-level saturation here, so
it has a higher potential for desynchronization than $n_b=2$,
which can be observed in the data in Fig.~\ref{fig:ChebFD_Spin}.
There is no prominent advantage of explicit overlap (\textsc{Pipeline}).
In fact, the speedup from removing the barrier synchronization
in the non-split version grows with
increasing communication volume along the strong scaling curve and at
certain points it becomes competitive with the \textsc{Pipeline} version.
Also, the barrier-free \textsc{Non-split} variant is consistently worse
for $n_b=2$, as expected.\smallskip

\highlight{\emph{Key takeaway}: Overlapping via explicit programming
  techniques may not be necessary for strongly bandwidth-saturating
  code with large (but not dominant) communication overhead due
  to the presence of  natural overlap by desynchronization.  }
		
\section{Outlook and future work}\label{sec:conclusion}

Using MPI-parallel synthetic benchmarks and application programs we
investigated the consequences of desynchronization via bottlenecks in
bulk-synchronous parallel code. Using a memory-bound microbenchmark we
showed that there is a strong positive correlation between idle wave
speed and the slope of a computational wave, indicating that automatic
communication-computation overlap can be more effective in settings
with low idle wave speeds. Using one-off idle injections we also
showed that a stable computational wave can be shifted along the MPI
rank dimension without losing its basic properties.

For back-to-back sparse matrix-vector multiplications, we demonstrated
that speedup by automatic desynchronization is facilitated by large
communication overhead, slow idle wave speed, and a simple non-split
communication scheme. Our investigation of a Chebyshev filter
diagonalization application showed that a more compact communication
topology, which can be affected by domain decomposition strategies
that allow for slower idle waves, enables more effective
communication overlap. Depending on the underlying problem (and thus
the sparse matrix structure), forcing overlap by explicit programming
may not even be required. 

We took great care to separate the effects of overhead reduction
via elimination of collective communication from the actual
benefit of desynchronization. Overall, our results show that
bottleneck evasion by desynchronization can be regarded
as a performance optimization technique, and that forcing
a parallel program into lock-step may be the wrong course
of action in some settings. 
%
%
	
\paragraph*{\textbf{Future work}}
In this work we have only considered memory bandwidth as the relevant
bottleneck in desynchronization phenomena, but we have reason
to assume that other bottlenecks, such as the compute node network injection
bandwidth or the network topology can effect similar
behavior in parallel codes. In future work we will explore this
option further. In addition we have as yet no rigorous proof
of instability for bottleneck-bound programs, which is why we
work towards an analytic description of desynchronization
processes that goes beyond idle wave speed.
	
In order to study out-of-lockstep behavior in more detail,
we are working on a message passing and threading simulator that
can simulate large-scale applications while taking the socket-level
properties of code into account. It can explore parallel program 
dynamics further in a controlled environment, saving
resources and time on real systems and allowing for
advanced architectural exploration. 

\appendices

\ifblind
\else
	\ifCLASSOPTIONcompsoc
	\section*{Acknowledgments}
	\else
	\section*{Acknowledgment}
	\fi
	
	This work was supported by KONWIHR, the Bavarian Competence Network
	for Scientific High Performance Computing in Bavaria, under the project name ``OMI4papps.''
	We are indebted to LRZ Garching for granting CPU hours on SuperMUC-NG.
	We wish to thank Andreas Alvermann for his ScaMaC library and the admin team at NHR@FAU for excellent technical support on Meggie system.
	
	\ifCLASSOPTIONcaptionsoff
	\newpage
	\fi
\fi


\bibliographystyle{IEEEtran}
\bibliography{acmart}

\begin{thebibliography}{10}
\providecommand{\url}[1]{#1}
\csname url@samestyle\endcsname
\providecommand{\newblock}{\relax}
\providecommand{\bibinfo}[2]{#2}
\providecommand{\BIBentrySTDinterwordspacing}{\spaceskip=0pt\relax}
\providecommand{\BIBentryALTinterwordstretchfactor}{4}
\providecommand{\BIBentryALTinterwordspacing}{\spaceskip=\fontdimen2\font plus
\BIBentryALTinterwordstretchfactor\fontdimen3\font minus
  \fontdimen4\font\relax}
\providecommand{\BIBforeignlanguage}[2]{{%
\expandafter\ifx\csname l@#1\endcsname\relax
\typeout{** WARNING: IEEEtran.bst: No hyphenation pattern has been}%
\typeout{** loaded for the language `#1'. Using the pattern for}%
\typeout{** the default language instead.}%
\else
\language=\csname l@#1\endcsname
\fi
#2}}
\providecommand{\BIBdecl}{\relax}
\BIBdecl

\bibitem{mccalpin1995memory}
J.~D. McCalpin \emph{et~al.}, ``Memory bandwidth and machine balance in current
  high performance computers,'' \emph{IEEE computer society technical committee
  on computer architecture (TCCA) newsletter}, vol.~2, no. 19--25, 1995.

\bibitem{pieper2016high}
A.~Pieper, M.~Kreutzer, A.~Alvermann, M.~Galgon, H.~Fehske, G.~Hager, B.~Lang,
  and G.~Wellein, ``High-performance implementation of {Chebyshev} filter
  diagonalization for interior eigenvalue computations,'' \emph{Journal of
  Computational Physics}, vol. 325, pp. 226--243, 2016.

\bibitem{carson2015communication}
E.~C. Carson, ``Communication-avoiding krylov subspace methods in theory and
  practice,'' Ph.D. dissertation, UC Berkeley, PZ, Italy, 2015.

\bibitem{GHYSELS2014224}
\BIBentryALTinterwordspacing
P.~Ghysels and W.~Vanroose, ``Hiding global synchronization latency in the
  preconditioned {Conjugate} {Gradient} algorithm,'' \emph{Parallel Computing},
  vol.~40, no.~7, pp. 224--238, 2014, 7th Workshop on Parallel Matrix
  Algorithms and Applications. [Online]. Available:
  \url{https://www.sciencedirect.com/science/article/pii/S0167819113000719}
\BIBentrySTDinterwordspacing

\bibitem{AfzalHW20}
A.~Afzal, G.~Hager, and G.~Wellein, ``{Desynchronization} and {Wave} {Pattern}
  {Formation} in {MPI}-{Parallel} and {Hybrid} {Memory}-{Bound} {Programs},''
  in \emph{Lecture Notes in Computer Science (including subseries Lecture Notes
  in Artificial Intelligence and Lecture Notes in Bioinformatics)},
  P.~Sadayappan, B.~L. Chamberlain, G.~Juckeland, and H.~Ltaief, Eds., vol.
  12151 LNCS.\hskip 1em plus 0.5em minus 0.4em\relax Cham: Springer
  International Publishing, 2020, pp. 391--411, cRIS-Team Scopus
  Importer:2020-07-10.

\bibitem{AfzalHW19}
------, ``{Propagation} and {Decay} of {Injected} {One}-{Off} {Delays} on
  {Clusters}: {A} {Case} {Study},'' in \emph{Proceedings - IEEE International
  Conference on Cluster Computing, ICCC}, vol. 2019-September.\hskip 1em plus
  0.5em minus 0.4em\relax Institute of Electrical and Electronics Engineers
  Inc., 2019, cRIS-Team Scopus Importer:2019-11-29.

\bibitem{Kreutzer:2015}
M.~{Kreutzer}, A.~{Pieper}, G.~{Hager}, G.~{Wellein}, A.~{Alvermann}, and
  H.~{Fehske}, ``Performance engineering of the {Kernel} {Polynomial} {Method}
  on large-scale {CPU}-{GPU} systems,'' in \emph{2015 IEEE International
  Parallel and Distributed Processing Symposium}, May 2015, pp. 417--426.

\bibitem{kreutzer2018chebyshev}
M.~Kreutzer, D.~Ernst, A.~R. Bishop, H.~Fehske, G.~Hager, K.~Nakajima, and
  G.~Wellein, ``Chebyshev filter diagonalization on modern manycore processors
  and {GPGPUs},'' in \emph{High Performance Computing}, R.~Yokota, M.~Weiland,
  D.~Keyes, and C.~Trinitis, Eds.\hskip 1em plus 0.5em minus 0.4em\relax Cham:
  Springer International Publishing, 2018, pp. 329--349.

\bibitem{AfzalHW2021}
A.~Afzal, G.~Hager, and G.~Wellein, ``{Analytic} {Modeling} of {Idle} {Waves}
  in {Parallel} {Programs}: {Communication}, {Cluster} {Topology}, and {Noise}
  {Impact},'' in \emph{Lecture Notes in Computer Science (including subseries
  Lecture Notes in Artificial Intelligence and Lecture Notes in
  Bioinformatics)}, B.~L. Chamberlain, A.-L. Varbanescu, H.~Ltaief, and
  P.~Luszczek, Eds., vol. 12728 LNCS.\hskip 1em plus 0.5em minus 0.4em\relax
  Springer Science and Business Media Deutschland GmbH, 2021, pp. 351--371,
  cRIS-Team Scopus Importer:2021-08-20.

\bibitem{AfzalHWcpe22}
\BIBentryALTinterwordspacing
------, ``{Analytic} performance model for parallel overlapping memory-bound
  kernels,'' \emph{Concurrency and Computation: Practice and Experience}, Jan
  2022. [Online]. Available:
  \url{https://onlinelibrary.wiley.com/doi/10.1002/cpe.6816}
\BIBentrySTDinterwordspacing

\bibitem{zhao2020hsm}
X.~Zhao, M.~Jahre, and L.~Eeckhout, ``Hsm: A hybrid slowdown model for
  multitasking gpus,'' in \emph{Proceedings of the Twenty-Fifth International
  Conference on Architectural Support for Programming Languages and Operating
  Systems}.\hskip 1em plus 0.5em minus 0.4em\relax ACM, 2020, pp. 1371--1385.

\bibitem{AfzalEuroMPI19Poster}
\BIBentryALTinterwordspacing
A.~Afzal, G.~Hager, and G.~Wellein, ``Delay flow mechanisms on clusters,''
  poster at EuroMPI 2019, September 10--13, 2019, Zurich, Switzerland.
  [Online]. Available:
  \url{https://hpc.fau.de/files/2019/09/EuroMPI2019_AHW-Poster.pdf}
\BIBentrySTDinterwordspacing

\bibitem{petrini2003case}
F.~Petrini, D.~J. Kerbyson, and S.~Pakin, ``The case of the missing
  supercomputer performance: {A}chieving optimal performance on the 8,192
  processors of {ASCI} {Q},'' in \emph{Supercomputing, 2003 ACM/IEEE
  Conference}.\hskip 1em plus 0.5em minus 0.4em\relax IEEE, 2003, pp. 55--55.

\bibitem{jones2003improving}
T.~Jones, S.~Dawson, R.~Neely, W.~Tuel, L.~Brenner, J.~Fier, R.~Blackmore,
  P.~Caffrey, B.~Maskell, P.~Tomlinson \emph{et~al.}, ``Improving the
  scalability of parallel jobs by adding parallel awareness to the operating
  system,'' in \emph{Supercomputing, 2003 ACM/IEEE Conference}.\hskip 1em plus
  0.5em minus 0.4em\relax IEEE, 2003, pp. 10--10.

\bibitem{terry2004improving}
P.~Terry, A.~Shan, and P.~Huttunen, ``Improving application performance on
  {HPC} systems with process synchronization.'' \emph{Linux Journal}, no. 127,
  pp. 68--71, 2004.

\bibitem{gioiosa2004analysis}
R.~Gioiosa, F.~Petrini, K.~Davis, and F.~Lebaillif-Delamare, ``Analysis of
  system overhead on parallel computers,'' in \emph{Proceedings of the Fourth
  IEEE International Symposium on Signal Processing and Information
  Technology}.\hskip 1em plus 0.5em minus 0.4em\relax IEEE, 2004, pp. 387--390.

\bibitem{tsafrir2005system}
D.~Tsafrir, Y.~Etsion, D.~G. Feitelson, and S.~Kirkpatrick, ``System noise,
  {OS} clock ticks, and fine-grained parallel applications,'' in
  \emph{Proceedings of the 19th annual international conference on
  Supercomputing}.\hskip 1em plus 0.5em minus 0.4em\relax ACM, 2005, pp.
  303--312.

\bibitem{beckman2006influence}
P.~Beckman, K.~Iskra, K.~Yoshii, and S.~Coghlan, ``The influence of operating
  systems on the performance of collective operations at extreme scale,'' in
  \emph{Cluster Computing, 2006 IEEE International Conference on}.\hskip 1em
  plus 0.5em minus 0.4em\relax IEEE, 2006, pp. 1--12.

\bibitem{ferreira2008characterizing}
K.~B. Ferreira, P.~Bridges, and R.~Brightwell, ``Characterizing application
  sensitivity to {OS} interference using kernel-level noise injection,'' in
  \emph{Proceedings of the 2008 ACM/IEEE conference on Supercomputing}.\hskip
  1em plus 0.5em minus 0.4em\relax IEEE Press, 2008, p.~19.

\bibitem{morari2011quantitative}
A.~Morari, R.~Gioiosa, R.~W. Wisniewski, F.~J. Cazorla, and M.~Valero, ``A
  quantitative analysis of {OS} noise,'' in \emph{2011 IEEE International
  Parallel \& Distributed Processing Symposium}.\hskip 1em plus 0.5em minus
  0.4em\relax IEEE, 2011, pp. 852--863.

\bibitem{Weisbach:2018}
H.~Weisbach, B.~Gerofi, B.~Kocoloski, H.~H{\"a}rtig, and Y.~Ishikawa,
  ``Hardware performance variation: A comparative study using lightweight
  kernels,'' in \emph{High Performance Computing}, R.~Yokota, M.~Weiland,
  D.~Keyes, and C.~Trinitis, Eds.\hskip 1em plus 0.5em minus 0.4em\relax Cham:
  Springer International Publishing, 2018, pp. 246--265.

\bibitem{hoefler2010characterizing}
T.~Hoefler, T.~Schneider, and A.~Lumsdaine, ``Characterizing the influence of
  system noise on large-scale applications by simulation,'' in
  \emph{Proceedings of the 2010 ACM/IEEE International Conference for High
  Performance Computing, Networking, Storage and Analysis}.\hskip 1em plus
  0.5em minus 0.4em\relax IEEE Computer Society, 2010, pp. 1--11.

\bibitem{markidis2015idle}
S.~Markidis, J.~Vencels, I.~B. Peng, D.~Akhmetova, E.~Laure, and P.~Henri,
  ``Idle waves in high-performance computing,'' \emph{Physical Review E},
  vol.~91, no.~1, p. 013306, 2015.

\bibitem{Gamell:2015}
M.~{Gamell}, K.~{Teranishi}, M.~A. {Heroux}, J.~{Mayo}, H.~{Kolla}, J.~{Chen},
  and M.~{Parashar}, ``Local recovery and failure masking for stencil-based
  applications at extreme scales,'' in \emph{SC '15: Proceedings of the
  International Conference for High Performance Computing, Networking, Storage
  and Analysis}, Nov 2015, pp. 1--12.

\bibitem{Boehme:2016}
D.~B{\"o}hme, M.~Geimer, L.~Arnold, F.~Voigtlaender, and F.~Wolf, ``Identifying
  the root causes of wait states in large-scale parallel applications,''
  \emph{ACM Trans. Parallel Comput.}, vol.~3, no.~2, pp. 11:1--11:24, Jul.
  2016.

\bibitem{kolakowska2004desynchronization}
A.~Kolakowska and M.~A. Novotny, ``Desynchronization and speedup in an
  asynchronous conservative parallel update protocol,'' \emph{arXiv preprint
  cs/0409032}, 2004.

\bibitem{AfzalISC21Poster}
\BIBentryALTinterwordspacing
A.~Afzal, G.~Hager, and G.~Wellein, ``{Physical} {Oscillator} {Model} for
  {Parallel} {Distributed} {Computing},'' 2021, poster at ISC High Performance
  2021. [Online]. Available:
  \url{https://www.researchgate.net/publication/354208484_Physical_Oscillator_Model_for_Parallel_Distributed_Computing?_sg=lVYEm3XW4E92lsf55nIWTxkXYhgpB13cA2Zi3zbsaP-YcPn2zMsmYHnIURLA_eADEZioHef0aYKgIpaCYNIPew0H4GtDTB14Q-E_avYu.W3TTE5Nzo7hjNZ45wagExeFeJB8qTBJQe764hOdV9pK0lbGaFk1muIggkEpWAveETapO_wOS8AKyPVJ1-4g0WA}
\BIBentrySTDinterwordspacing

\bibitem{an2011score}
D.~an~Mey, S.~Biersdorf, C.~Bischof, K.~Diethelm, D.~Eschweiler, M.~Gerndt,
  A.~Kn{\"u}pfer, D.~Lorenz, A.~Malony, W.~E. Nagel \emph{et~al.}, ``Score-p: A
  unified performance measurement system for petascale applications,'' in
  \emph{Competence in High Performance Computing 2010}.\hskip 1em plus 0.5em
  minus 0.4em\relax Springer, 2011, pp. 85--97.

\bibitem{Neural2019}
T.-T. Pham, M.~Pister, and P.~Couvée, ``Recurrent neural network for
  classifying of hpc applications,'' in \emph{2019 Spring Simulation Conference
  (SpringSim)}, 2019, pp. 1--12.

\bibitem{hager2010introduction}
G.~Hager and G.~Wellein, \emph{Introduction to high performance computing for
  scientists and engineers}.\hskip 1em plus 0.5em minus 0.4em\relax CRC Press,
  2010.

\bibitem{gropp1999toward}
W.~D. Gropp, D.~K. Kaushik, D.~E. Keyes, and B.~F. Smith, ``Toward realistic
  performance bounds for implicit cfd codes,'' in \emph{Proceedings of parallel
  CFD}, vol.~99.\hskip 1em plus 0.5em minus 0.4em\relax Citeseer, 1999, pp.
  233--240.

\bibitem{Moritz2014}
\BIBentryALTinterwordspacing
M.~Kreutzer, G.~Hager, G.~Wellein, H.~Fehske, and A.~R. Bishop, ``{A} unified
  sparse matrix data format for efficient general sparse matrix-vector
  multiplication on modern processors with wide {SIMD} units,'' \emph{SIAM
  Journal on Scientific Computing}, vol.~36, pp. C401--C423, 2014. [Online].
  Available: \url{http://epubs.siam.org/doi/abs/10.1137/130930352}
\BIBentrySTDinterwordspacing

\bibitem{fehske2004quantum}
H.~Fehske, G.~Wellein, G.~Hager, A.~Wei{\ss}e, and A.~Bishop, ``Quantum lattice
  dynamical effects on single-particle excitations in one-dimensional mott and
  peierls insulators,'' \emph{Physical Review B}, vol.~69, no.~16, p. 165115,
  2004.

\bibitem{GHOST2017}
M.~Kreutzer, J.~Thies, M.~R{\"o}hrig-Z{\"o}llner, A.~Pieper, F.~Shahzad,
  M.~Galgon, A.~Basermann, H.~Fehske, G.~Hager, and G.~Wellein, ``{GHOST}:
  Building blocks for high performance sparse linear algebra on heterogeneous
  systems,'' \emph{International Journal of Parallel Programming}, vol.~45, pp.
  1046--1072, 2017.

\bibitem{sitte2012topological}
M.~Sitte, A.~Rosch, E.~Altman, and L.~Fritz, ``Topological insulators in
  magnetic fields: {Quantum} {Hall} effect and edge channels with a
  nonquantized theta term,'' \emph{Physical review letters}, vol. 108, no.~12,
  p. 126807, 2012.

\end{thebibliography}

\begin{IEEEbiography}[{\includegraphics[width=1.05in,height=1.9in,clip,keepaspectratio]{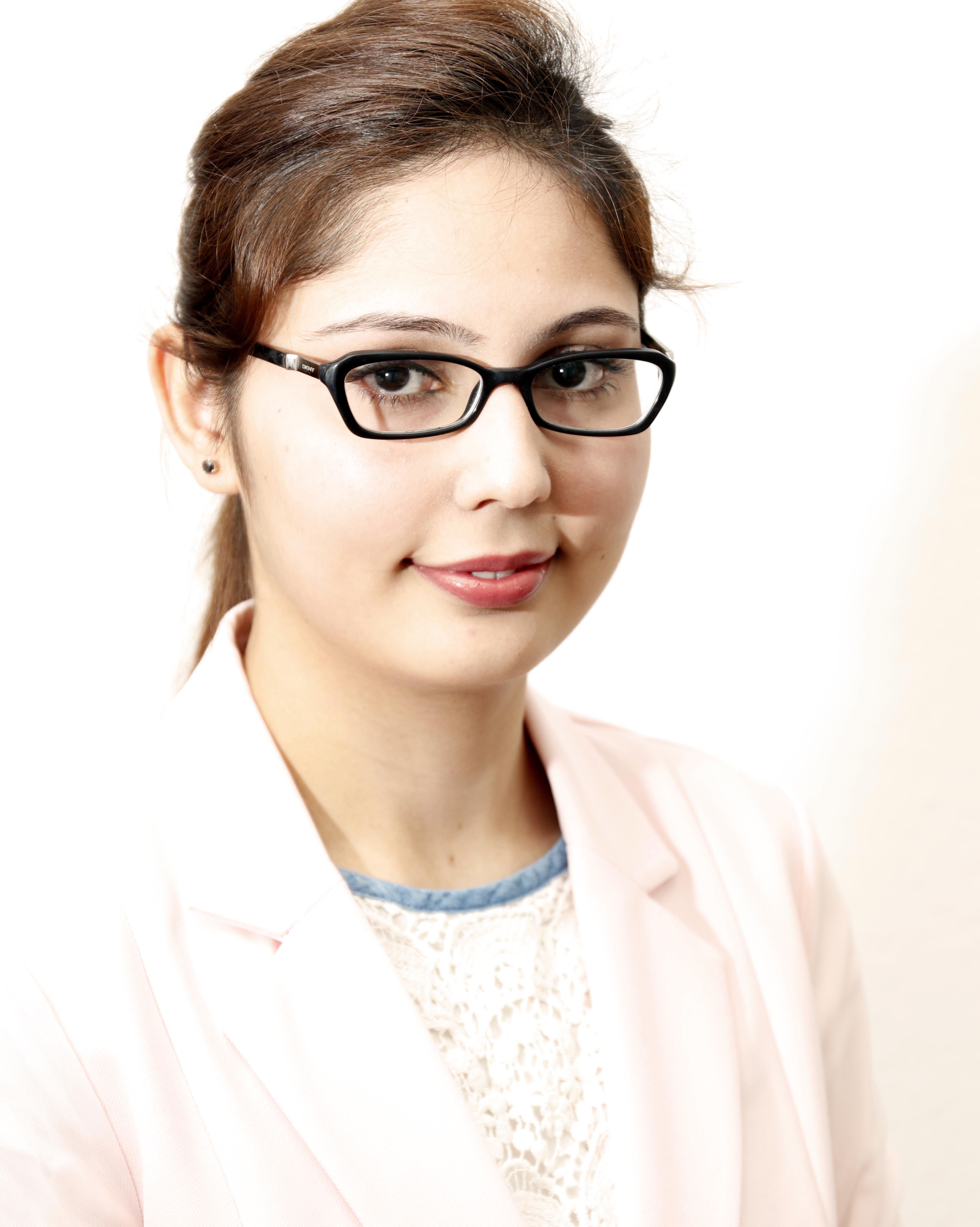}}]{Ayesha Afzal}
	holds a Master’s degree in Computational Engineering from Friedrich-Alexander-Universit\"at Erlangen-N\"urnberg, Germany, followed by a Bachelor’s degree in Electrical Engineering from the University of Engineering and Technology, Lahore, Pakistan.
	She is working toward the Ph.D. degree at the professorship for High Performance Computing at Erlangen National High Performance Computing Center (NHR@FAU), Germany.
	Her PhD research lies at the intersection of analytic performance models, performance tools and parallel simulation frameworks, with a focus on first-principles performance modeling of distributed-memory parallel programs in high-performance computing.
\end{IEEEbiography}
\begin{IEEEbiography}[{\includegraphics[width=1.05in,height=1.9in,clip,keepaspectratio]{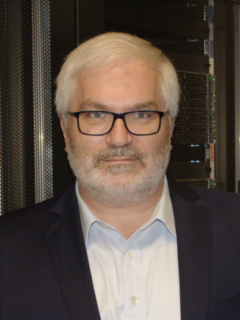}}]{Georg Hager} holds a doctorate (Ph.D.) and a Habilitation degree in Computational Physics from the University of Greifs\-wald, Germany. He leads the Training \& Support Division at Erlangen National High Performance Computing Center (NHR@FAU) and is an associate lecturer at the Institute of Physics at the University of Greifs\-wald. Recent research includes architecture-specific optimization strategies for current microprocessors, performance engineering of scientific codes on chip and system levels, and the modeling of out-of-lockstep behavior in large-scale parallel codes.
\end{IEEEbiography}
\begin{IEEEbiography}[{\includegraphics[width=1.05in,height=1.9in,clip,keepaspectratio]{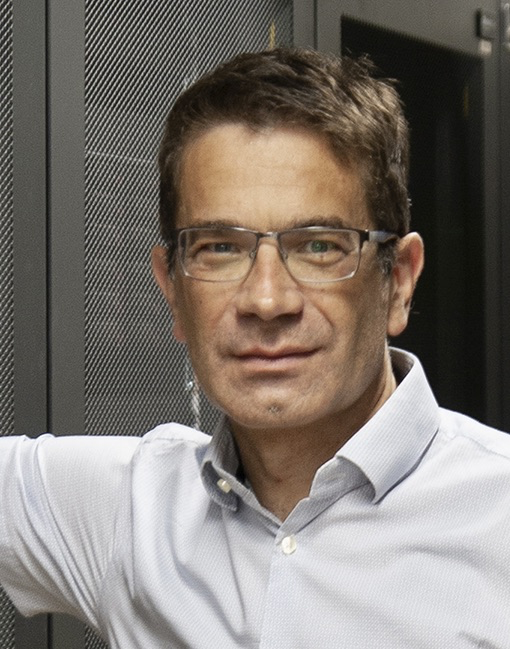}}]{Gerhard Wellein}
 received the Diploma (M.Sc.) degree and a doctorate (Ph.D.) degree in Physics from the University of Bayreuth, Germany. He is a Professor at the Department of Computer Science at Friedrich-Alexander-Universit\"at Erlangen-N\"urnberg and heads the Erlangen National Center for High-Performance Computing (NHR@FAU). His research interests focus on performance modeling and performance engineering, architecture-specific code optimization, and hardware-efficient building blocks for sparse linear algebra and stencil solvers.
\end{IEEEbiography}

\end{document}